\documentclass[%
 reprint,
 amsmath,amssymb,citeautoscript,
 aip,
]{revtex4-2}

\usepackage{graphicx}
\usepackage{dcolumn}
\usepackage{natbib}
\usepackage{bm}

\newcommand{\kB}{k_{\mathrm{B}}}

\usepackage{chemist}
\usepackage[version=4]{mhchem}

\begin{document}



\title{Enhanced ClNO$_2$ formation at the interface of sea-salt aerosol}

\author{Seokjin Moon}
 \affiliation{Department of Chemistry, University of California, Berkeley, CA, USA}
\author{David T. Limmer}
 \email{dlimmer@berkeley.edu}
\affiliation{Department of Chemistry, University of California, Berkeley, CA, USA}
\affiliation{Kavli Energy NanoScience Institute, Berkeley, CA, USA}
\affiliation{\mbox{Materials Sciences Division, Lawrence Berkeley National Laboratory, Berkeley, CA, USA}}
\affiliation{\mbox{Chemical Sciences Division, Lawrence Berkeley National Laboratory, Berkeley, CA, USA}}

\date{\today}

\begin{abstract}
The reactive uptake of $\mathrm{N_2O_5}$ on sea-spray aerosol plays a key role in regulating NO$_\mathrm{x}$ concentration in the troposphere. Despite numerous field and laboratory studies, a microscopic understanding of its heterogeneous reactivity remains unclear. Here, we use molecular simulation and theory to elucidate the chlorination of $\mathrm{N_2O_5}$ to form ClNO$_2$, the primary reactive channel within sea-spray aerosol. We find the formation of ClNO$_2$ is markedly enhanced at the air-water interface due to the stabilization of the charge-delocalized transition state, as evident from the formulation of bimolecular rate theory in heterogeneous environments. We explore the consequences of the enhanced interfacial reactivity in the uptake of $\mathrm{N_2O_5}$ using numerical solutions of molecular reaction-diffusion equations as well as their analytical approximations. Our results suggest that the current interpretation of aerosol branching ratios needs to be revisited.
\end{abstract}
\maketitle


\section{\label{sec:introduction}Introduction}

When trace gas molecules collide with an aerosol particle, a mass accommodation process transports them into the bulk of the aerosol where some are lost irreversibly to subsequent chemical reactions. The efficiency of this heterogeneous reactive uptake plays a crucial role in global-scale environmental processes\cite{abbatt2012quantifying, tie2001effects}. The uptake efficiency depends on elementary steps---including adsorption, desorption, diffusion and reaction---whose relative importance in heterogeneous environments is challenging to anticipate\cite{limmer2024molecular}. Surface-specific spectroscopy\cite{piatkowski2014extreme, litman2024surface} and molecular simulations\cite{galib2021reactive, cruzeiro2022uptake} have revealed that physical and chemical properties of a molecularly-thin region around the air-water interface are distinct from either of the surrounding bulk phases, complicating studies of heterogeneous uptake. We have developed a  molecular model of this complex interface to understand the uptake of $\mathrm{N_2O_5}$ to form ClNO$_2$ on the surface of a sea-salt aerosol.

The uptake of $\mathrm{N_2O_5}$ by aerosol has been investigated to understand the fate of tropospheric $\mathrm{NO_x}$ at night\cite{davis2008parameterization, brown2012nighttime}. Tropospheric $\mathrm{NO_x}$ radicals are the primary source of hydroxyl radical and catalyze ozone production\cite{dentener1993reaction}. A global chemistry transport model estimated 41\% of $\mathrm{NO_x}$ are consumed through the heterogeneous reactive uptake of $\mathrm{N_2O_5}$\cite{alexander2020global}. The key parameter for the global model is the reactive uptake coefficient, $\gamma$, a ratio of the irreversibly lost reactive flux to the kinetic surface collision flux. Interpretation of the measured variation of $\gamma$ in terms of aerosol composition and meteorological conditions\cite{bertram2009toward, stewart2004reactive, gaston2016reacto} demands a  microscopic perspective, which ultimately would enable the prediction and parameterization of $\gamma$. For sea-spray aerosols, arguably the two most important reactions determining the uptake of $\mathrm{N_2O_5}$ are
\begin{align}
    \mathrm{{N_2}{O_5}_{(g)} + H_2O_{(l)}} &\reactrarrow{0pt}{1.0cm}{$k_\mathrm{H}$}{} \mathrm{2HN{O_3}_{(aq)}} \tag{R1} \\
    \mathrm{{N_2}{O_5}_{(g)} + Cl^-_{(aq)}} &\reactrarrow{0pt}{1.0cm}{$k_\mathrm{C}$}{} \mathrm{ClN{O_2}_{(aq)} + N{O_3}^-_{(aq)} }\tag{R2} 
\end{align}
hydrolysis to nitric acid (R1) and chlorination to $\mathrm{ClNO_2}$ (R2). Atomistic simulations based on machine-learning forcefields have provided thermodynamic and mechanistic detail into the hydrolysis reaction (R1)\cite{cruzeiro2022uptake, galib2021reactive}. However, a complementary microscopic understanding of the chlorination reaction (R2) is lacking\cite{kercher2009chlorine}. 

Early studies of (R2) with cryogenic vibrational spectroscopy observed a stable ion-dipole complex, $[\mathrm{ClN_2O_5}]^-$, illustrating the plausibility of an $\mathrm{S_N2}$-type reaction mechanism\cite{kelleher2017trapping}. Subsequent first-principles calculations have found low-lying reaction pathways in water clusters consistent with this perspective\cite{karimova2019sn2, mccaslin2023effects}. For an $\mathrm{S_N2}$-type charge transfer reaction with $\mathrm{Cl^-}$, solvent molecules around the solute need to be collectively rearranged to accommodate the distinct charge distribution of the product. As a consequence, the role of the heterogeneous environment cannot be ignored for this reaction. Molecular dynamics simulations with a reactive model we developed based on the empirical valence bond approach\cite{warshel1991computer} suggest that the chlorination rate is significantly enhanced in the vicinity of the air-water interface. The important role of the solvent degrees of freedom emerged naturally from analysis of the commitment probability\cite{hummer2004transition}.

Understanding how an enhanced reaction rate at the air-water interface contributes to $\gamma$ requires a mass transport equation. Continuum reaction-diffusion equations are widely used\cite{davidovits2006mass}, which describe diffusive dynamics of various chemical species with their reactivity. However, most are phenomenological in that bulk transport properties are used for their parameterization\cite{danckwerts1951absorption}. We formulate the molecularly-detailed reaction-diffusion equation for $\mathrm{N_2O_5}$ with two reaction channels, (R1) and (R2). Numerical solutions with various bulk $\mathrm{Cl^-}$ concentrations are compared with experimental measurements, and challenge the previous analytical framework based on bulk-dominant and homogeneous reactivity. Moreover, we propose a new, analytic, approximate solution to the reaction-diffusion equation. The novel expression, together with the molecular simulations, paves the way to systematically incorporate heterogeneous reactivity to the analysis of the reactive uptake process. 

\section*{\label{sec:propensity}Surface propensities of solutes}
\begin{figure}[t]
    \centering
    \includegraphics[width=0.5\textwidth]{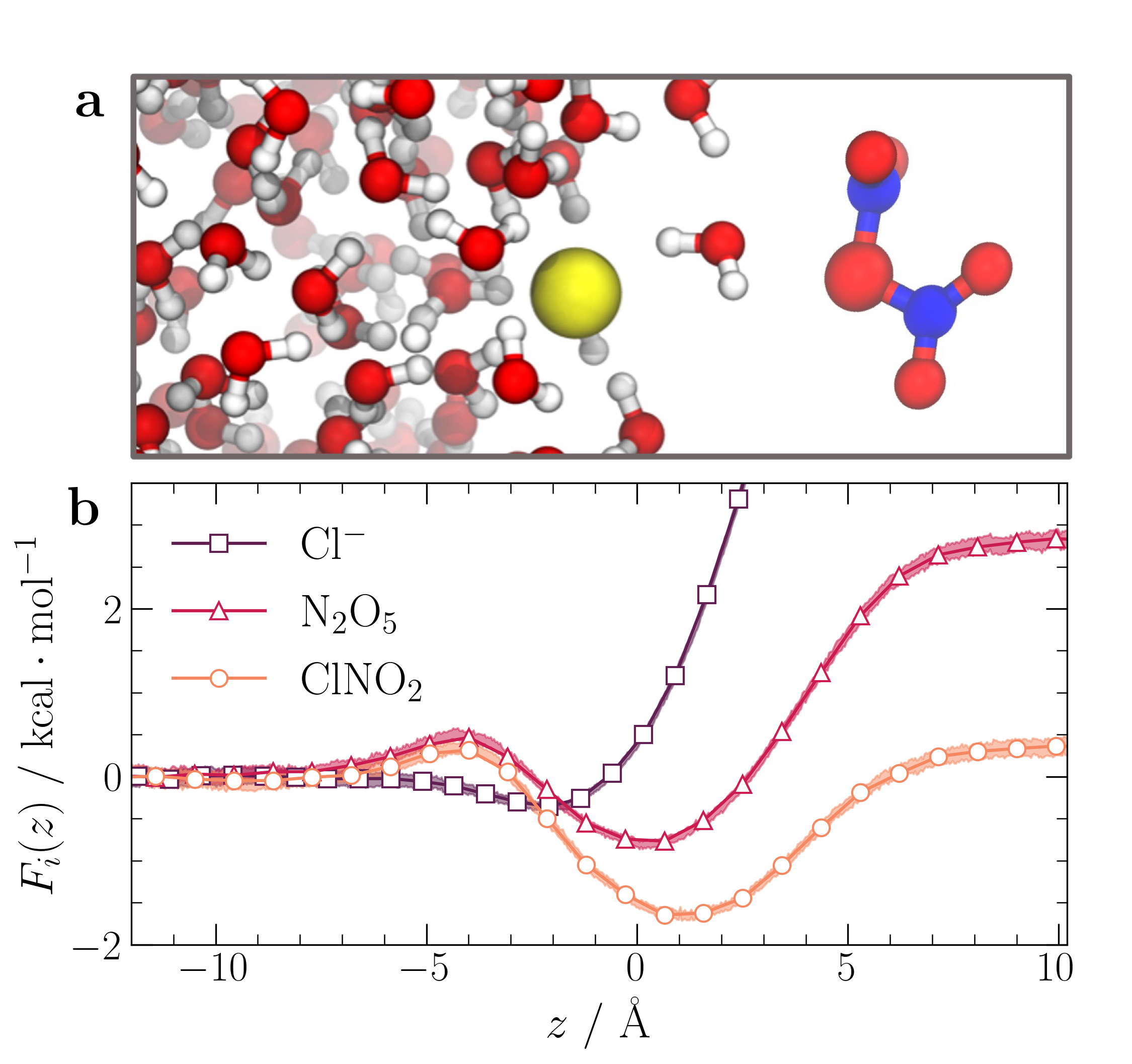}
    \caption{(a) Schematic picture of the air-water interface system with chemicals in (R2). (b) Reversible work as a function of center-of-mass $z$ location for each species $i=\mathrm{Cl^{-}}$, N$_2$O$_5$, and ClNO$_2$. Gibbs dividing surface is located at $z=0\AA$. Positive and negative $z$ correspond to the bulk gas and solution sides, respectively. Errors represent 1 standard error. }
    \label{fig:1}
\end{figure}

The air-water interface is a molecularly thin environment where the solvent behaves in a manner distinct from its bulk aqueous phase. The reduced amount of solvent at the interface results in an altered thermodynamic cost to rearrange the hydrogen-bonding structure, which influences affinities of reactants to the interface as well as the rate of reactions driven by electrostatic fluctuations near it\cite{ballard2012toward,kattirtzi2017microscopic,geissler2001autoionization, reischl2009statistics}. To provide a balanced description of both reactivity as well as the long-ranged electrostatic environment of an explicit air-water interface, we developed a novel empirical valence bond model. We built our model by constructing two low-lying diabatic states, encoding the reactant and product states of (R2). The diabatic Hamiltonian for each state was re-parameterized from the generalized Amber force field\cite{wang2004development}. The point charges on each atom were optimized to reproduce the interfacial affinities of each species estimated from  higher levels of theory\cite{cruzeiro2022uptake, ou2013spherical}. The two diabatic states were coupled through a diabatic coupling Hamiltonian modeled as a linear combination of multiple Gaussians\cite{schlegel2006empirical} and fitted to reproduce the gas-phase potential energy surface computed with the range--separated hybrid density functional, $\omega$B97X--V/DEF2--TZVPD\cite{mardirossian2014omegab97x}. The water was modeled using the flexible variant of SPC/Fw\cite{wu2006flexible}. Further details are provided in supplementary information. With the reactive model, we simulated the air-water interface using a slab geometry consisting of 512 water molecules, a single $\mathrm{N_2O_5}$, and one excess Cl$^-$. A schematic picture of the system is shown in Fig.~\ref{fig:1}a.
 
The reversible work for moving a molecule perpendicular to the interface measures its interfacial propensity. We evaluate the reversible work, or the free energy profile, for each chemical species in (R2) as a function of the depth from the interface, $F_i(z)=-\kB T \ln \langle \delta (z-z_c^i)\rangle$, where $z_c^i$ is the center of mass of species $i$ relative to the Gibbs dividing surface, $\kB T$ is Boltzmann's constant times temperature, and the brackets surrounding the Dirac delta function denotes a canonical ensemble average at fixed temperature $298\mathrm{K}$. Umbrella sampling with harmonic biasing potentials was applied to evaluate these averages\cite{frenkel2023understanding} and histograms of $z$ from such simulations were re-weighted using the weighted histogram analysis method\cite{kumar1995multidimensional} to compute an unbiased estimate of $F_i(z)$. 

The resultant free energy profiles for Cl$^-$, $\mathrm{N_2O_5}$, and ClNO$_2$ are shown in Fig.~\ref{fig:1}b. The two neutral molecules, $\mathrm{N_2O_5}$ and $\mathrm{ClNO_2}$, are locally stabilized around the air-water interface, and both face free energy barriers to move from the interface into either bulk vapor or bulk solution phases.This is due to the weak interactions with surrounding water molecules, resulting in an enriched concentration of both species at the interface relative to the bulk. By contrast, strong ion-water interactions prevent $\mathrm{Cl^-}$ from penetrating into the vapor\cite{loche2022molecular}, which is evident from the stiff rise of the free energy starting at $z=0\mathrm{\AA}$. A slight stabilization of $\mathrm{Cl^-}$ around the interface is consistent with some surface-specific spectroscopy measurements\cite{piatkowski2014extreme}.

The free energy difference between the gas and aqueous phases is related to a well-known descriptor of solubility known as Henry's law constant. Direct measurements of the Henry's law constant for $\mathrm{N_2O_5}$ are limited due to its facile hydrolysis. Indirect measurements report a value of $H=2.1\mathrm{M/atm}$\cite{fried1994reaction}, which is close to our estimate of $H=4.8\mathrm{M/atm}$. For $\mathrm{ClNO_2}$, lab measurements report $H=0.024\mathrm{M/atm}$\cite{behnke1997production} and we estimate $H=0.040\mathrm{M/atm}$. A lower value of $H$ indicates more volatile nature of $\mathrm{ClNO_2}$ compare to $\mathrm{N_2O_5}$. Considered in conjunction with the expected exothermic forward reaction with $\mathrm{Cl^-}$\cite{karimova2019sn2, mccaslin2023effects}, the role of the backward reaction of (R2) is considered to be negligible in the uptake process. 

\begin{figure*}[ht]
    \centering
    \includegraphics[width=\textwidth]{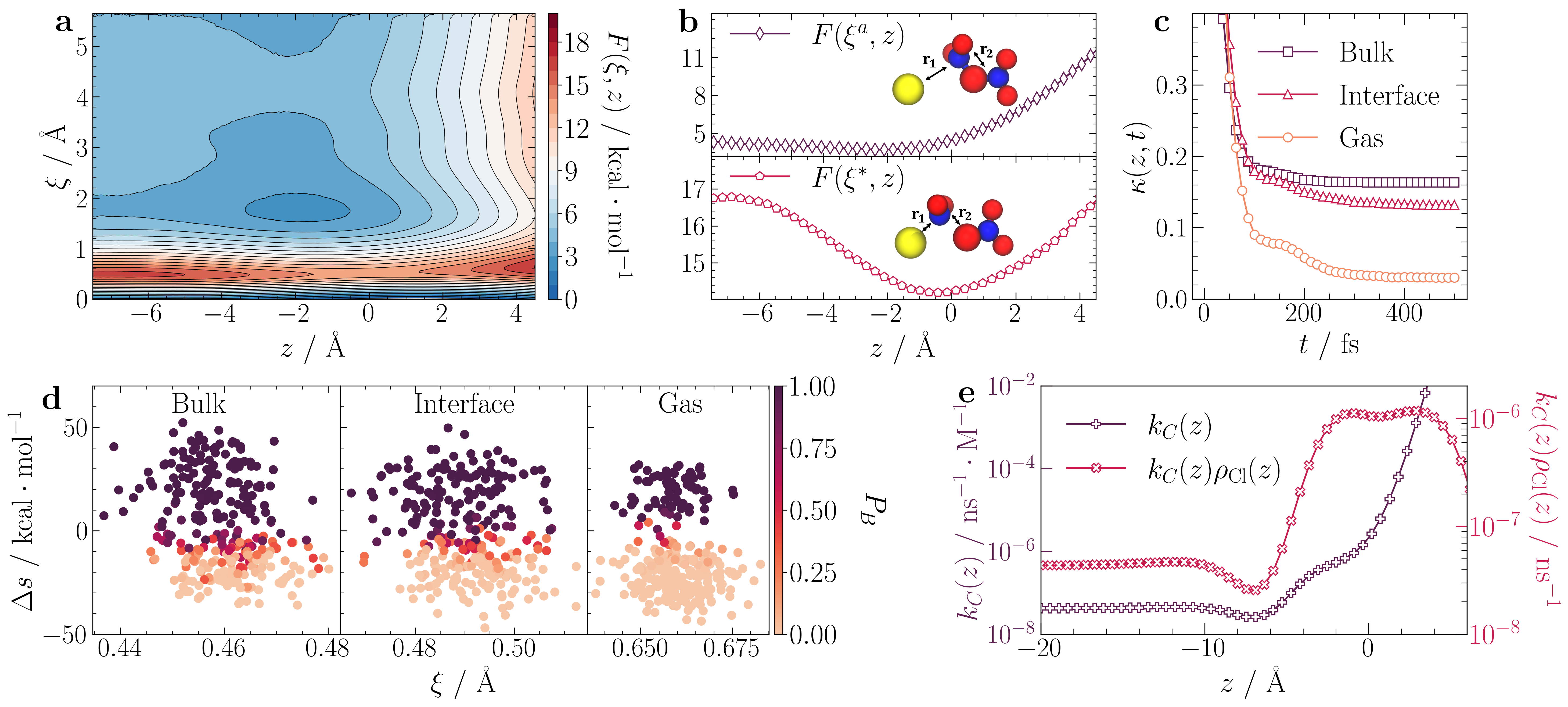}
    \caption{(a) Free energy surface as a function of $\mathrm{S_N2}$ order parameter $\xi$ and depth $z$ of $\mathrm{Cl^-}$. (b) Conditional free energy profile along $z$ for the reactant ($\xi^a=5.62\mathrm{\AA}$) and transition state  ($\xi^*=0.52\mathrm{\AA}$), with schematic pictures of states.
    (c) Transmission coefficients measured at three distinct $z$ constraints. (d) Commitment probability as a function of $\xi$ and solvents' degrees of freedom $\Delta s$ at three $z$ values.  (e) Heterogeneous bimolecular rate (left y-axis) and pseudo first-order loss rate of $\mathrm{N_2O_5}$ (right y-axis) corresponding to $1 \mathrm{M}$ of bulk $\mathrm{Cl^-}$ density. For (c) and (d), the $z$ values correspond to Bulk ($z=-12.3\mathrm{\AA}$), Interface ($z=0\mathrm{\AA}$) and Gas ($z=5\mathrm{\AA}$) }
    \label{fig:2}
\end{figure*}

\section*{\label{sec:pmf}Bimolecular rate enhancement}
The substitution reaction in (R2) is expected to be concerted. A canonical order parameter for describing it would be $\xi=r_1-r_2$, which is defined as a bond-length difference between bond-breaking and bond-forming pairs. In this case, the leaving group is the nitrate anion, and so we take $r_2$ to be the distance between the center oxygen on N$_2$O$_5$ and the nitrogen forming ClNO$_2$. The nucleophile is Cl$^-$, and so we take $r_1$ to be the distance between it and the same nitrogen. While $\xi$ is sufficient to differentiate the reactant and product states in a bulk solution, it does not discriminate between the reaction occurring at or away from the air-water interface. In order to clarify the thermodynamics of the reaction and how it depends in proximity to the interface, we consider the free energy surface defined by both $\xi$ and the $z$ position of $\mathrm{Cl^-}$. The resultant free energy, $F(\xi,z)=- \kB T \ln \langle \delta(z-z^\mathrm{Cl}_c)\delta (\xi-r_1+r_2)\rangle$ is computed in an analogous manner to its marginal $F_{\mathrm{Cl}}(z)$ using umbrella sampling and histogram reweighting.

Figure \ref{fig:2}a depicts $F(\xi,z)$ in the vicinity of the Gibbs dividing surface. Three states are evident along constant $z$ slices: a reactant state where $\mathrm{Cl^{-1}}$ is distant from $\mathrm{N_2O_5}$ with $\xi = 5.62\mathrm{\AA}$, an ion-dipole complex for $\xi \sim 2 \mathrm{\AA}$, and a transition state whose free energies are local maximum occurring near $\xi \sim 0.5\mathrm{\AA}$. The free energetics of the reaction share common features across $z$;the ion-dipole complex is metastable and the reaction is strongly exothermic. The ion-dipole complex is most metastable at the interface due to the interplay of incomplete dielectric screening and partial solvation. The stabilization of the reactant and the barrier height also show variations near the interface. This is illustrated in Fig.~\ref{fig:2}b, where constant $\xi$ slices of the free energy surface are shown; $F(\xi^a,z)$ for the reactant state and $F(\xi^*,z)$ for the transition state. The $z$ dependence of the reactant free energy mirrors the stability of the Cl$^-$, which is thermodynamically unfavorable to desolvate. The $z$ dependence of free energy at the transition state is non-monotonic, exhibiting a minimum at the Gibbs dividing surface. The lowering of the barrier height from the bulk solution to the interface is significant, which is nearly $3\mathrm{kcal\cdot mol^{-1}}$. 

The transition state of the $\mathrm{S_N2}$ reaction described by our model has a delocalized charge distribution, with half of the negative charge transferred to the nitrate ion.  Our model potential energy surface in the gas phase exhibits a sub-$\mathrm{kcal\cdot mol^{-1}}$ barrier due to the facile ability of $\mathrm{N_2O_5}$ to dissociate. Hence, the gas-phase reaction is expected to proceed at the diffusion limit, like most ion-dipole chemistry. In the bulk solution phase, the thermodynamic penalty for distributing the charge across the Cl and NO$_3$ species results in a significant free energy barrier. The destabilization results from the relatively unfavorable solvation free energy of the delocalized charge distribution and is the origin of the slow reaction kinetics in polar solvents that have been observed experimentally\cite{hamlin2018nucleophilic} and theoretically\cite{benjamin2008empirical}. The existence of the minimum in the barrier height at the interface  reflects the competing factors affecting the solvation of charged species at the air-water interface\cite{shaffer2013ion,jungwirth2001molecular}. The charge-delocalized transition state behaves similarly to large anions that have shown preferential adsorption due to a combination of favorable excluded volume factors arising because of facile density fluctuations and unfavorable electrostatic contributions such as repulsive image charge interactions\cite{otten2012elucidating}.

Connecting the free energetics of the reaction to its rate requires a rate theory. To describe the process in (R2), we need to generalize the standard bimolecular rate theory for a heterogeneous environment. Detailed discussion on the derivation of such a generalization can be found in the supplementary information, but it follows from manipulating the usual side-side correlation function and imposing a conditioning on the reaction occurring at a fixed $z$ position\cite{schile2019rate}. Within a reactive flux formulation, the bimolecular rate can be written as a product of a position dependent transmission coefficient $\kappa(z)$ and a transition state theory (TST) approximation\cite{laidler1983development} to the rate, $k_\mathrm{C}^\mathrm{TST}(z)$
\begin{align}
    k^\mathrm{TST}_\mathrm{C}(z) = \left[ \sqrt{\frac{\kB T|\mathbf{G}|}{2\pi}}  \frac{3}{r_1^a} e^{- \Delta F(z)/\kB T} \right] \nu(z) \, , \label{eqn:heterorate}
\end{align}
where $\Delta F(z)=F(\xi^*, z)-F(\xi^a, z)$ is the barrier height at fixed $z$ and $|\mathbf{G}|$ is the inverse effective mass of $\xi$\cite{berne1988classical}. The factor $\nu(z)$ accounts for the local density enhancement of $\mathrm{N_2O_5}$ and is given by,
\begin{align}
    \nu(z) =  \int_{0}^{r_1^a} r^2 dr \int d\Omega \, e^{- [F_\mathrm{N}(z+r cos(\theta))-F_\mathrm{N}(z)]/\kB T} \, , \label{eqn:Lz}
\end{align}
where $d\Omega = \sin(\theta) d\theta d\phi$ are polar angles oriented along $z$, and $F_\mathrm{N}(z)$ is the free energy profile of $\mathrm{N_2O_5}$ shown in Fig.~\ref{fig:1}b. The local $\mathrm{N_2O_5}$ density is averaged over a size given as $r_1^a$, determined by where the free energy obtains an asymptotic scaling, $F(\xi,z) = F(\xi^a,z) - 2 \kB T \ln \left(r_1/r_1^{a} \right)$ and $\xi$ is well approximated by $r_1$, i.e. $\xi^a\approx r_1^a$. The expression incorporates two contributions. The part in the squared bracket accounts for the barrier crossing frequency if $\mathrm{N_2O_5}$ is constrained in the sphere of radius $r_1^a$ centered on $\mathrm{Cl^-}$. This frequency depends on the free energy barrier to the transition state from the reactant state, not from the ion-dipole state. This is because of the unbounded domain of the bimolecular order parameter, $\xi$, which results in overwhelming probability of being in the separated state compared to the bound state. The factor $\nu(z)$ accounts for the ratio of the mean density of $\mathrm{N_2O_5}$ inside the sphere to the local density at the $z$ location.

The transmission coefficient, $\kappa(z)$, provides a correction to the transition state approximation of the rate by accounting for recrossing events. Using Bennett-Chandler approach\cite{chandler1978statistical}, we measured it using the time-correlation function, $\kappa(z,t) =2 \left< v(0) h[\xi(t)] \right>^{*}_{z}/\left<|v|\right>$, where $v$ is a velocity of $\xi$, and $h$ is an indicator function which has a value of 1 if $\xi(t)$ is in the product domain or 0 otherwise. The average is taken over the ensemble of trajectories initialized from the transition state with specific $z$ value. The asymptotic long-time limit of $\kappa(z, t)$ in Fig.~\ref{fig:2}c is the correction factor $\kappa(z)$, which converged to 0.15 in bulk aqueous environment, 0.14 at the interface, and 0.045 $\mathrm{\AA}$ above the interface. Hence, the dynamic correction to the rate is insensitive to the environment unless the reaction occurs far from the condensed phase.

The small values of $\kappa(z)$ computed suggest that $\xi$ does not fully describe the relevant reaction coordinate of the transition. To uncover the mechanism, we performed a committor analysis by introducing another order parameter $\Delta s$, the solvent's contribution to the reorganization energy. In detail, $\Delta s$ is the result of subtracting the solvent-solute interaction energy of the product diabatic state from that of the reactant diabatic state\cite{benjamin2008empirical}. The commitment probability, $\mathrm{P_B}(\xi, \Delta s)$, measures the relative number of trajectories started from fixed $\xi$ and $\Delta s$, reaching the product state before the reactant state\cite{hummer2004transition}. As such, $P_B$ quantifies the order parameters' performance, since it is 0.5 if the fixed initial point is a member of the transition state ensemble. As shown in Fig.~\ref{fig:2}d, the iso-commitment points are aligned near zero of $\Delta s$, indicating the important role of solvent's degrees of freedom. Fluctuations of $\Delta s$ at a constrained value of $\xi$ result in the small values of $\kappa$ computed, and its insensitivity with $z$ demonstrates that the mechanism of a solvent mediated charge transfer is conserved near and from the interface.

The bimolecular rate is given by $k_\mathrm{C}(z) = \kappa(z) k_\mathrm{C}^\mathrm{TST}(z)$, and is shown in Fig~\ref{fig:2}e. We observe a markedly enhanced reaction rate at the interface,  increasing below the Gibbs dividing surface, and steeply rising above it. The rate depends exponentially on the the free energy difference between two states in Fig.~\ref{fig:2}b. Thus, the rate enhancement below the interface is mainly attributed to the stability of the transition state, while its steep rise above it is due to the destabilization of $\mathrm{Cl^-}$. This becomes more apparent when the equilibrium density profile of $\mathrm{Cl^-}$ is multiplied by the bimolecular rate. The density profile can be computed from a bulk density $\bar{\rho}_\mathrm{Cl}$ times the marginal free energy for Cl$^-$, $\rho_\mathrm{Cl}(z)=\bar{\rho}_\mathrm{Cl} \exp[-F_\mathrm{Cl}(z)/\kB T]$. The pseudo first-order rate, $k_\mathrm{C}(z)\rho_\mathrm{Cl}(z)$, with $\bar{\rho}_\mathrm{Cl}=1\mathrm{M}$ is also shown in Fig.~\ref{fig:2}e, which shows clear local peak in the rate by 25 times around the interface relative to the bulk solution. The subsequent decrease in the rate at large $z$ results from the thermodynamic difficulty of getting the transition state into the gas phase. The variation in the rate occurs across molecular scales, and is constant a nanometer away from the interface.

\section*{\label{sec:kinetics}Molecularly resolved kinetic uptake model}
The heterogeneous reactive uptake of $\mathrm{N_2O_5}$ involves a series of physical and chemical processes coupled to the diffusive motion. Hence, a coupled partial differential equation for each process is required to be solved to estimate the reactive uptake coefficient. While analytical solutions are only given for some limiting cases\cite{danckwerts1970gas}, approximate solutions based on the decoupling of the processes are provided as the resistor model\cite{worsnop1989temperature}. The resistor model has been widely used as a reasonable framework to analyze measured variations of trace gas uptake coefficients\cite{davidovits2006mass}. However, given the significant variation in the substitution rate with position relative to the interface we have found, a molecularly-detailed reaction-diffusion model is necessary.

We employ an overdamped Fokker-Planck equation for the density of $\mathrm{N_2O_5}$, parameterized from our molecular dynamics simulations\cite{polley2024statistical}. The time-dependent density profile of $\mathrm{N_2O_5}$, $\rho_\mathrm{N}(z, t)$, satisfies
\begin{align}
    &\partial_t \rho_\mathrm{N} = \partial_z D(z) e^{-\beta F_\mathrm{N}(z)} \partial_z e^{\beta F_\mathrm{N}(z)} \rho_\mathrm{N}- k_\mathrm{T}(z) \rho_\mathrm{N} \, , \label{eqn:RDE} 
\end{align}
where $D(z)$ is the spatially dependent diffusion coefficient and $k_\mathrm{T}(z)=k_\mathrm{H}(z)+k_\mathrm{C}(z)\rho_\mathrm{Cl}(z)$ is the spatially dependent total rate for two reaction channels. Solving Eq.~\ref{eqn:RDE} with an absorbing boundary condition placed at $z=6.58\mathrm{\AA}$, we are able to model the evaporation process without introducing underdamped dynamics. We also use a reflecting boundary condition located at $z= -200\mathrm{nm}$ to model a semi-infinite slab of solution. Propagating the equation from a normalized Gaussian-shaped density profile centered on the Gibbs dividing surface describes a distribution of molecules thermalized at the interface after initial collisions with unit sticking probability. The uptake coefficient $\gamma$ is measured from the solution of Eq.~\ref{eqn:RDE} as a loss of density due to the reaction,
\begin{align}
    \gamma_\mathrm{T} = \int_{0}^{\infty} dt \int dz \, k_\mathrm{T}(z) \rho_\mathrm{N}(z, t) \, , \label{eqn:uptake}
\end{align}
which in this model competes only with re-evaporation.

\begin{figure}[ht]
    \centering
    \includegraphics[width=0.49\textwidth]{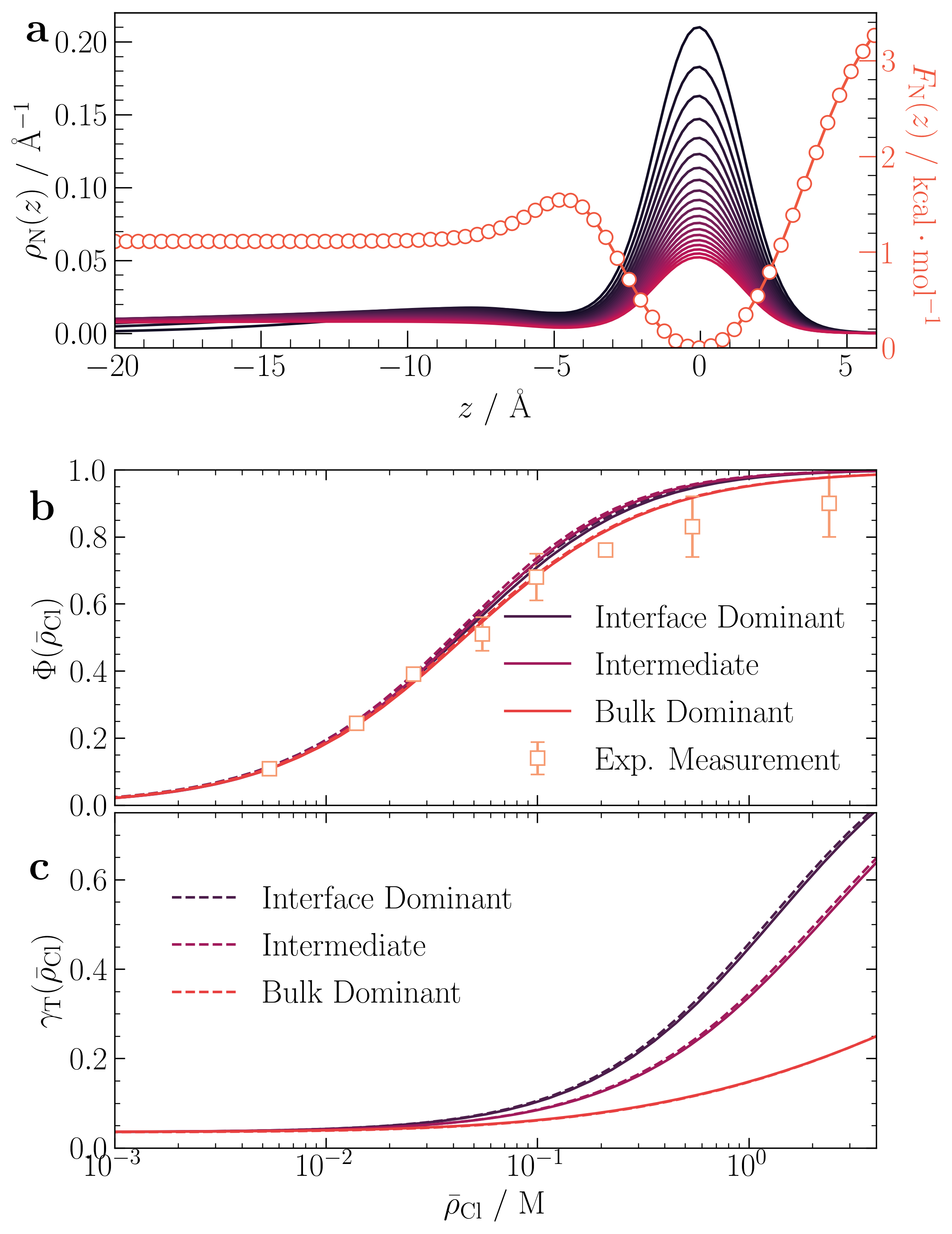}
    \caption{(a)Time-dependent density profiles of $\mathrm{N_2O_5}$ with $\bar{\rho}_\mathrm{Cl}=0$.  Curves are separated by 18 ns. The free energy profile of $\mathrm{N_2O_5}$ is shown on the right-axis. (b) The $\mathrm{ClNO_2}$ branching ratio measured from Eq.~\ref{eqn:branching} (solid lines) and the analytic formula (dashed line) as a function of $\bar{\rho}_\mathrm{Cl}$. Flow reactor measurements from Kregel. et. al. are shown as square markers.\cite{kregel2023weak} (c) Reactive uptake coefficients from Eq.~\ref{eqn:uptake} (solid lines) and the analytic formula (dashed line) as a function of $\bar{\rho}_\mathrm{Cl}$. }
    \label{fig:3}
\end{figure}

Using a step function form for $D(z)$ employed previously with a slightly enhanced interfacial diffusivity relative to the bulk solution\cite{cruzeiro2022uptake}, we first considered the uptake due to just hydrolysis, $\bar{\rho}_\mathrm{Cl}=0\mathrm{M}$. We found $k_\mathrm{H} = 0.4/\mu s$ results in the experimentally reported value of $\gamma = 0.036$ for a pure water droplet\cite{gaston2016reacto}. This rate is much slower than previous theoretical estimates in the literature\cite{galib2021reactive, cruzeiro2022uptake}, while close to experimental estimates\cite{gaston2016reacto}, reflecting the slower evaporation rate estimated with our model. Under these conditions the impact of the spatial dependence of $k_\mathrm{H}(z)$ was minor, and so we have treated it as a constant. The resulting density profiles obtained by solving Eq.~\ref{eqn:RDE} with finite differences using the optimized form of $k_\mathrm{H}(z)$ are shown in Fig.~\ref{fig:3}a. The initial Gaussian distribution relaxes over a 100 nanoseconds and begins to mirror the equilibrium density determined by $F_\mathrm{N}(z)$, though with reduced amplitude due to evaporation and hydrolysis.

Next, the hydrolysis and chlorination reactions are considered together. In this two reaction channel problem, the branching ratio or the $\mathrm{ClNO_2}$ production yield, $\Phi$, becomes a key measurement for analyzing reactive uptake process. Within our reaction diffusion model, it can be computed from
\begin{align}
    \Phi =\frac{1}{\gamma_\mathrm{T}}\int_{0}^{\infty} dt \int dz \, k_\mathrm{C}(z) \rho_\mathrm{Cl}(z) \rho_\mathrm{N}(z, t) \, , \label{eqn:branching}
\end{align}
which is a ratio of the reactive flux into the chlorination channel relative to the total loss due to reactions. Recent experimental measurements have reported $\Phi$ as a function of bulk Cl$^-$ concentration, and are shown in Fig.~\ref{fig:3}b\cite{kregel2023weak}. We estimated $\Phi$ using the rate profile $k_{\mathrm{C}}(z)$ from Fig.~\ref{fig:2}c under pseudo-first order conditions. However, the bare rate predicted from our model is far too slow. Given the gas phase energetics comparable to high level CCSD(T) calculations in Fig.~\ref{fig:S2}, the aqueous reaction barrier is overestimated due to overly stabilizing water-solute interactions. If we assume that the shape of rate profile is robust since the enhancement originated from the augmented propensity of charge-delocalized transition state at the interface, then we can include a multiplicative factor to the rate profile to reproduce the experimental branching ratio of $\Phi=0.108$ with $\bar{\rho}_\mathrm{Cl}=0.0054 \mathrm{M}$, and probe the behavior of the system away from that point. 
 
The branching ratio and uptake coefficients using this procedure are summarized in Fig.~\ref{fig:3}b and \ref{fig:3}c labeled ``Intermediate". We have also considered two additional hypothetical scenarios, an ``Interface Dominant" case, whose rate is enhanced by 250 times at the interface and ``Bulk Dominant" case with zero interfacial rate. Both rate profiles are shifted in the same way as ``Intermediate" case to reproduce $\Phi$ at $\bar{\rho}_\mathrm{Cl}=0.0054$ M. Interestingly the estimated branching ratios show minor differences among the scenarios. However, heterogeneous reactivity has a pronounced effect on the $\bar{\rho}_\mathrm{Cl}$-dependent $\gamma$. The total uptake coefficient at moderately concentrated solution distinguishes these three clearly. In each case, the rise in $\Phi$ occurs at a lower Cl$^-$ concentration then the subsequent rise in $\gamma$. 

Previous analysis of $\mathrm{N_2O_5}$ reactive uptake in the chloride-containing aerosol has relied on the framework of resistor model\cite{bertram2009toward, kregel2023weak}. Within that framework, the rates and concentrations are assumed to be uniform across the solution, and the branching ratio is subsequently interpreted as a ratio of bulk rate constants. While it cannot be applied directly to our molecularly resolved model, we have found a minimal modification of the resistor model that is capable of reproducing the trends. We consider the total uptake as $\gamma=\gamma_{\mathrm{I}}+\gamma_{\mathrm{B}}(1-\gamma_{\mathrm{I}})$. The $\gamma_{\mathrm{I}}$ is an ideal interfacial uptake, 
\begin{align}
    \gamma_\mathrm{I}= \frac{ k_\mathrm{I,H} + k_\mathrm{I,C}\bar{\rho}_\mathrm{Cl} }{k_e + k_\mathrm{I,H} + k_\mathrm{I,C}\bar{\rho}_\mathrm{Cl}} \, ,
\end{align}
where $k_\mathrm{I,H}$ and $k_\mathrm{I,C}$ are interfacial rates for hydrolysis and chlorination, respectively, and $k_e$ is the evaporation rate. The bulk uptake
\begin{align}
    \gamma_\mathrm{B}= \left (\frac{1}{\alpha'}+\frac{\bar{v}}{4H' \sqrt{D(k_\mathrm{B,H}+ k_\mathrm{B,C}\bar{\rho}_\mathrm{Cl})}} \right )^{-1}
\end{align}
depends on a renormalized mass accommodation coefficient $\alpha'$ and Henry's law constant $H'$,
\begin{align}
     \alpha' = \frac{k_{s}}{k_{s} + k_{e} + k_\mathrm{I,H} + k_\mathrm{I,C}\bar{\rho}_\mathrm{Cl} }, \indent H' = H (1-\gamma_\mathrm{I}) \, ,
\end{align}
where $D$ is the bulk diffusivity, $\bar{v}$ is molecular thermal velocity, $k_\mathrm{B,H}$ and $k_\mathrm{B,C}$ are bulk rates for hydrolysis and chlorination, respectively, and $k_{s}$ is the solvation rate. The dashed lines in Fig.~\ref{fig:3}b and \ref{fig:3}c are results from the new expression, which reproduce the numerical results remarkably well. A detailed derivation can be found in the supporting material, but this expression reduces to known forms in the limit that the uptake is determined fully by either the bulk or interfacial processes.

Our molecularly-detailed reaction-diffusion equation emphasizes the importance of spatial heterogeneity. Rate profiles whose interfacial contribution vary over 2 orders of magnitude produce the same observable branching ratio, which complicate the traditional interpretation of the branching ratio. For example, weak temperature dependence of the branching ratio was attributed to similar activation barriers for two reactions\cite{kregel2023weak}. However, activation barriers at the interface may differ from the bulk reaction and so may need to be considered together. While measurements of $\Phi$ have provided valuable insights into the competitive nature of two reaction channels, it is clear that the available information is insufficient to conclusively determine the significance of interfacial contribution to the overall reactive uptake process. A simultaneous measurement of the total uptake and the branching ratio from experiments is necessary. More broadly, the observation here that a S$_\mathrm{N}2$ type reaction can be markedly accelerated at a liquid-vapor interface due to the stabilization of a charge delocalized transition state suggests a potential origin for recent observations of chemistry in microdroplets and \emph{on-water} catalysis\cite{qiu2021reaction}. The framework employed here using a combination of molecular modeling and theory, can be straightforwardly translated into those cases as well. 

\begin{acknowledgments}
The authors would like to thank Gil Nathanson for extensive discussions. D.T.L was supported during the early part of this work by  the National Science Foundation through the
National Science Foundation Center for Aerosol Impacts on Chemistry of the Environment (NSF-CAICE) under grant number CHE 1801971.  D.T.L. acknowledges support from an Alfred P. Sloan Research Fellowship. S. M. was supported by NSF Grant
CHE2102314.
\end{acknowledgments}

\section*{References}

\nocite{*}

\bibliography{Reference.bib}

\begin{thebibliography}{61}%
\makeatletter
\providecommand \@ifxundefined [1]{%
 \@ifx{#1\undefined}
}%
\providecommand \@ifnum [1]{%
 \ifnum #1\expandafter \@firstoftwo
 \else \expandafter \@secondoftwo
 \fi
}%
\providecommand \@ifx [1]{%
 \ifx #1\expandafter \@firstoftwo
 \else \expandafter \@secondoftwo
 \fi
}%
\providecommand \natexlab [1]{#1}%
\providecommand \enquote  [1]{``#1''}%
\providecommand \bibnamefont  [1]{#1}%
\providecommand \bibfnamefont [1]{#1}%
\providecommand \citenamefont [1]{#1}%
\providecommand \href@noop [0]{\@secondoftwo}%
\providecommand \href [0]{\begingroup \@sanitize@url \@href}%
\providecommand \@href[1]{\@@startlink{#1}\@@href}%
\providecommand \@@href[1]{\endgroup#1\@@endlink}%
\providecommand \@sanitize@url [0]{\catcode `\\12\catcode `\$12\catcode
  `\&12\catcode `\#12\catcode `\^12\catcode `\_12\catcode `\%12\relax}%
\providecommand \@@startlink[1]{}%
\providecommand \@@endlink[0]{}%
\providecommand \url  [0]{\begingroup\@sanitize@url \@url }%
\providecommand \@url [1]{\endgroup\@href {#1}{\urlprefix }}%
\providecommand \urlprefix  [0]{URL }%
\providecommand \Eprint [0]{\href }%
\providecommand \doibase [0]{https://doi.org/}%
\providecommand \selectlanguage [0]{\@gobble}%
\providecommand \bibinfo  [0]{\@secondoftwo}%
\providecommand \bibfield  [0]{\@secondoftwo}%
\providecommand \translation [1]{[#1]}%
\providecommand \BibitemOpen [0]{}%
\providecommand \bibitemStop [0]{}%
\providecommand \bibitemNoStop [0]{.\EOS\space}%
\providecommand \EOS [0]{\spacefactor3000\relax}%
\providecommand \BibitemShut  [1]{\csname bibitem#1\endcsname}%
\let\auto@bib@innerbib\@empty
\bibitem [{\citenamefont {Abbatt}, \citenamefont {Lee},\ and\ \citenamefont
  {Thornton}(2012)}]{abbatt2012quantifying}%
  \BibitemOpen
  \bibfield  {author} {\bibinfo {author} {\bibfnamefont {J.}~\bibnamefont
  {Abbatt}}, \bibinfo {author} {\bibfnamefont {A.}~\bibnamefont {Lee}},\ and\
  \bibinfo {author} {\bibfnamefont {J.}~\bibnamefont {Thornton}},\ }\bibfield
  {title} {\enquote {\bibinfo {title} {Quantifying trace gas uptake to
  tropospheric aerosol: recent advances and remaining challenges},}\
  }\href@noop {} {\bibfield  {journal} {\bibinfo  {journal} {Chemical Society
  Reviews}\ }\textbf {\bibinfo {volume} {41}},\ \bibinfo {pages} {6555--6581}
  (\bibinfo {year} {2012})}\BibitemShut {NoStop}%
\bibitem [{\citenamefont {Tie}\ \emph {et~al.}(2001)\citenamefont {Tie},
  \citenamefont {Brasseur}, \citenamefont {Emmons}, \citenamefont {Horowitz},\
  and\ \citenamefont {Kinnison}}]{tie2001effects}%
  \BibitemOpen
  \bibfield  {author} {\bibinfo {author} {\bibfnamefont {X.}~\bibnamefont
  {Tie}}, \bibinfo {author} {\bibfnamefont {G.}~\bibnamefont {Brasseur}},
  \bibinfo {author} {\bibfnamefont {L.}~\bibnamefont {Emmons}}, \bibinfo
  {author} {\bibfnamefont {L.}~\bibnamefont {Horowitz}},\ and\ \bibinfo
  {author} {\bibfnamefont {D.}~\bibnamefont {Kinnison}},\ }\bibfield  {title}
  {\enquote {\bibinfo {title} {Effects of aerosols on tropospheric oxidants: A
  global model study},}\ }\href@noop {} {\bibfield  {journal} {\bibinfo
  {journal} {Journal of Geophysical Research: Atmospheres}\ }\textbf {\bibinfo
  {volume} {106}},\ \bibinfo {pages} {22931--22964} (\bibinfo {year}
  {2001})}\BibitemShut {NoStop}%
\bibitem [{\citenamefont {Limmer}\ \emph {et~al.}(2024)\citenamefont {Limmer},
  \citenamefont {G{\"o}tz}, \citenamefont {Bertram},\ and\ \citenamefont
  {Nathanson}}]{limmer2024molecular}%
  \BibitemOpen
  \bibfield  {author} {\bibinfo {author} {\bibfnamefont {D.~T.}\ \bibnamefont
  {Limmer}}, \bibinfo {author} {\bibfnamefont {A.~W.}\ \bibnamefont
  {G{\"o}tz}}, \bibinfo {author} {\bibfnamefont {T.~H.}\ \bibnamefont
  {Bertram}},\ and\ \bibinfo {author} {\bibfnamefont {G.~M.}\ \bibnamefont
  {Nathanson}},\ }\bibfield  {title} {\enquote {\bibinfo {title} {Molecular
  insights into chemical reactions at aqueous aerosol interfaces},}\
  }\href@noop {} {\bibfield  {journal} {\bibinfo  {journal} {Annual Review of
  Physical Chemistry}\ }\textbf {\bibinfo {volume} {75}} (\bibinfo {year}
  {2024})}\BibitemShut {NoStop}%
\bibitem [{\citenamefont {Piatkowski}\ \emph {et~al.}(2014)\citenamefont
  {Piatkowski}, \citenamefont {Zhang}, \citenamefont {Backus}, \citenamefont
  {Bakker},\ and\ \citenamefont {Bonn}}]{piatkowski2014extreme}%
  \BibitemOpen
  \bibfield  {author} {\bibinfo {author} {\bibfnamefont {L.}~\bibnamefont
  {Piatkowski}}, \bibinfo {author} {\bibfnamefont {Z.}~\bibnamefont {Zhang}},
  \bibinfo {author} {\bibfnamefont {E.~H.}\ \bibnamefont {Backus}}, \bibinfo
  {author} {\bibfnamefont {H.~J.}\ \bibnamefont {Bakker}},\ and\ \bibinfo
  {author} {\bibfnamefont {M.}~\bibnamefont {Bonn}},\ }\bibfield  {title}
  {\enquote {\bibinfo {title} {Extreme surface propensity of halide ions in
  water},}\ }\href@noop {} {\bibfield  {journal} {\bibinfo  {journal} {Nature
  communications}\ }\textbf {\bibinfo {volume} {5}},\ \bibinfo {pages} {4083}
  (\bibinfo {year} {2014})}\BibitemShut {NoStop}%
\bibitem [{\citenamefont {Litman}\ \emph {et~al.}(2024)\citenamefont {Litman},
  \citenamefont {Chiang}, \citenamefont {Seki}, \citenamefont {Nagata},\ and\
  \citenamefont {Bonn}}]{litman2024surface}%
  \BibitemOpen
  \bibfield  {author} {\bibinfo {author} {\bibfnamefont {Y.}~\bibnamefont
  {Litman}}, \bibinfo {author} {\bibfnamefont {K.-Y.}\ \bibnamefont {Chiang}},
  \bibinfo {author} {\bibfnamefont {T.}~\bibnamefont {Seki}}, \bibinfo {author}
  {\bibfnamefont {Y.}~\bibnamefont {Nagata}},\ and\ \bibinfo {author}
  {\bibfnamefont {M.}~\bibnamefont {Bonn}},\ }\bibfield  {title} {\enquote
  {\bibinfo {title} {Surface stratification determines the interfacial water
  structure of simple electrolyte solutions},}\ }\href@noop {} {\bibfield
  {journal} {\bibinfo  {journal} {Nature Chemistry}\ ,\ \bibinfo {pages}
  {1--7}} (\bibinfo {year} {2024})}\BibitemShut {NoStop}%
\bibitem [{\citenamefont {Galib}\ and\ \citenamefont
  {Limmer}(2021)}]{galib2021reactive}%
  \BibitemOpen
  \bibfield  {author} {\bibinfo {author} {\bibfnamefont {M.}~\bibnamefont
  {Galib}}\ and\ \bibinfo {author} {\bibfnamefont {D.~T.}\ \bibnamefont
  {Limmer}},\ }\bibfield  {title} {\enquote {\bibinfo {title} {Reactive uptake
  of n2o5 by atmospheric aerosol is dominated by interfacial processes},}\
  }\href@noop {} {\bibfield  {journal} {\bibinfo  {journal} {Science}\ }\textbf
  {\bibinfo {volume} {371}},\ \bibinfo {pages} {921--925} (\bibinfo {year}
  {2021})}\BibitemShut {NoStop}%
\bibitem [{\citenamefont {Cruzeiro}\ \emph {et~al.}(2022)\citenamefont
  {Cruzeiro}, \citenamefont {Galib}, \citenamefont {Limmer},\ and\
  \citenamefont {G{\"o}tz}}]{cruzeiro2022uptake}%
  \BibitemOpen
  \bibfield  {author} {\bibinfo {author} {\bibfnamefont {V.~W.~D.}\
  \bibnamefont {Cruzeiro}}, \bibinfo {author} {\bibfnamefont {M.}~\bibnamefont
  {Galib}}, \bibinfo {author} {\bibfnamefont {D.~T.}\ \bibnamefont {Limmer}},\
  and\ \bibinfo {author} {\bibfnamefont {A.~W.}\ \bibnamefont {G{\"o}tz}},\
  }\bibfield  {title} {\enquote {\bibinfo {title} {Uptake of n2o5 by aqueous
  aerosol unveiled using chemically accurate many-body potentials},}\
  }\href@noop {} {\bibfield  {journal} {\bibinfo  {journal} {Nature
  communications}\ }\textbf {\bibinfo {volume} {13}},\ \bibinfo {pages} {1266}
  (\bibinfo {year} {2022})}\BibitemShut {NoStop}%
\bibitem [{\citenamefont {Davis}, \citenamefont {Bhave},\ and\ \citenamefont
  {Foley}(2008)}]{davis2008parameterization}%
  \BibitemOpen
  \bibfield  {author} {\bibinfo {author} {\bibfnamefont {J.}~\bibnamefont
  {Davis}}, \bibinfo {author} {\bibfnamefont {P.}~\bibnamefont {Bhave}},\ and\
  \bibinfo {author} {\bibfnamefont {K.}~\bibnamefont {Foley}},\ }\bibfield
  {title} {\enquote {\bibinfo {title} {Parameterization of n 2 o 5 reaction
  probabilities on the surface of particles containing ammonium, sulfate, and
  nitrate},}\ }\href@noop {} {\bibfield  {journal} {\bibinfo  {journal}
  {Atmospheric Chemistry and Physics}\ }\textbf {\bibinfo {volume} {8}},\
  \bibinfo {pages} {5295--5311} (\bibinfo {year} {2008})}\BibitemShut {NoStop}%
\bibitem [{\citenamefont {Brown}\ and\ \citenamefont
  {Stutz}(2012)}]{brown2012nighttime}%
  \BibitemOpen
  \bibfield  {author} {\bibinfo {author} {\bibfnamefont {S.~S.}\ \bibnamefont
  {Brown}}\ and\ \bibinfo {author} {\bibfnamefont {J.}~\bibnamefont {Stutz}},\
  }\bibfield  {title} {\enquote {\bibinfo {title} {Nighttime radical
  observations and chemistry},}\ }\href@noop {} {\bibfield  {journal} {\bibinfo
   {journal} {Chemical Society Reviews}\ }\textbf {\bibinfo {volume} {41}},\
  \bibinfo {pages} {6405--6447} (\bibinfo {year} {2012})}\BibitemShut {NoStop}%
\bibitem [{\citenamefont {Dentener}\ and\ \citenamefont
  {Crutzen}(1993)}]{dentener1993reaction}%
  \BibitemOpen
  \bibfield  {author} {\bibinfo {author} {\bibfnamefont {F.~J.}\ \bibnamefont
  {Dentener}}\ and\ \bibinfo {author} {\bibfnamefont {P.~J.}\ \bibnamefont
  {Crutzen}},\ }\bibfield  {title} {\enquote {\bibinfo {title} {Reaction of
  n2o5 on tropospheric aerosols: Impact on the global distributions of no x,
  o3, and oh},}\ }\href@noop {} {\bibfield  {journal} {\bibinfo  {journal}
  {Journal of Geophysical Research: Atmospheres}\ }\textbf {\bibinfo {volume}
  {98}},\ \bibinfo {pages} {7149--7163} (\bibinfo {year} {1993})}\BibitemShut
  {NoStop}%
\bibitem [{\citenamefont {Alexander}\ \emph {et~al.}(2020)\citenamefont
  {Alexander}, \citenamefont {Sherwen}, \citenamefont {Holmes}, \citenamefont
  {Fisher}, \citenamefont {Chen}, \citenamefont {Evans},\ and\ \citenamefont
  {Kasibhatla}}]{alexander2020global}%
  \BibitemOpen
  \bibfield  {author} {\bibinfo {author} {\bibfnamefont {B.}~\bibnamefont
  {Alexander}}, \bibinfo {author} {\bibfnamefont {T.}~\bibnamefont {Sherwen}},
  \bibinfo {author} {\bibfnamefont {C.~D.}\ \bibnamefont {Holmes}}, \bibinfo
  {author} {\bibfnamefont {J.~A.}\ \bibnamefont {Fisher}}, \bibinfo {author}
  {\bibfnamefont {Q.}~\bibnamefont {Chen}}, \bibinfo {author} {\bibfnamefont
  {M.~J.}\ \bibnamefont {Evans}},\ and\ \bibinfo {author} {\bibfnamefont
  {P.}~\bibnamefont {Kasibhatla}},\ }\bibfield  {title} {\enquote {\bibinfo
  {title} {Global inorganic nitrate production mechanisms: comparison of a
  global model with nitrate isotope observations},}\ }\href@noop {} {\bibfield
  {journal} {\bibinfo  {journal} {Atmospheric Chemistry and Physics}\ }\textbf
  {\bibinfo {volume} {20}},\ \bibinfo {pages} {3859--3877} (\bibinfo {year}
  {2020})}\BibitemShut {NoStop}%
\bibitem [{\citenamefont {Bertram}\ and\ \citenamefont
  {Thornton}(2009)}]{bertram2009toward}%
  \BibitemOpen
  \bibfield  {author} {\bibinfo {author} {\bibfnamefont {T.}~\bibnamefont
  {Bertram}}\ and\ \bibinfo {author} {\bibfnamefont {J.}~\bibnamefont
  {Thornton}},\ }\bibfield  {title} {\enquote {\bibinfo {title} {Toward a
  general parameterization of n 2 o 5 reactivity on aqueous particles: the
  competing effects of particle liquid water, nitrate and chloride},}\
  }\href@noop {} {\bibfield  {journal} {\bibinfo  {journal} {Atmospheric
  Chemistry and Physics}\ }\textbf {\bibinfo {volume} {9}},\ \bibinfo {pages}
  {8351--8363} (\bibinfo {year} {2009})}\BibitemShut {NoStop}%
\bibitem [{\citenamefont {Stewart}, \citenamefont {Griffiths},\ and\
  \citenamefont {Cox}(2004)}]{stewart2004reactive}%
  \BibitemOpen
  \bibfield  {author} {\bibinfo {author} {\bibfnamefont {D.~J.}\ \bibnamefont
  {Stewart}}, \bibinfo {author} {\bibfnamefont {P.}~\bibnamefont {Griffiths}},\
  and\ \bibinfo {author} {\bibfnamefont {R.}~\bibnamefont {Cox}},\ }\bibfield
  {title} {\enquote {\bibinfo {title} {Reactive uptake coefficients for
  heterogeneous reaction of n 2 o 5 with submicron aerosols of nacl and natural
  sea salt},}\ }\href@noop {} {\bibfield  {journal} {\bibinfo  {journal}
  {Atmospheric Chemistry and Physics}\ }\textbf {\bibinfo {volume} {4}},\
  \bibinfo {pages} {1381--1388} (\bibinfo {year} {2004})}\BibitemShut {NoStop}%
\bibitem [{\citenamefont {Gaston}\ and\ \citenamefont
  {Thornton}(2016)}]{gaston2016reacto}%
  \BibitemOpen
  \bibfield  {author} {\bibinfo {author} {\bibfnamefont {C.~J.}\ \bibnamefont
  {Gaston}}\ and\ \bibinfo {author} {\bibfnamefont {J.~A.}\ \bibnamefont
  {Thornton}},\ }\bibfield  {title} {\enquote {\bibinfo {title}
  {Reacto-diffusive length of n2o5 in aqueous sulfate-and chloride-containing
  aerosol particles},}\ }\href@noop {} {\bibfield  {journal} {\bibinfo
  {journal} {The Journal of Physical Chemistry A}\ }\textbf {\bibinfo {volume}
  {120}},\ \bibinfo {pages} {1039--1045} (\bibinfo {year} {2016})}\BibitemShut
  {NoStop}%
\bibitem [{\citenamefont {Kercher}, \citenamefont {Riedel},\ and\ \citenamefont
  {Thornton}(2009)}]{kercher2009chlorine}%
  \BibitemOpen
  \bibfield  {author} {\bibinfo {author} {\bibfnamefont {J.}~\bibnamefont
  {Kercher}}, \bibinfo {author} {\bibfnamefont {T.}~\bibnamefont {Riedel}},\
  and\ \bibinfo {author} {\bibfnamefont {J.}~\bibnamefont {Thornton}},\
  }\bibfield  {title} {\enquote {\bibinfo {title} {Chlorine activation by n 2 o
  5: simultaneous, in situ detection of clno 2 and n 2 o 5 by chemical
  ionization mass spectrometry},}\ }\href@noop {} {\bibfield  {journal}
  {\bibinfo  {journal} {Atmospheric Measurement Techniques}\ }\textbf {\bibinfo
  {volume} {2}},\ \bibinfo {pages} {193--204} (\bibinfo {year}
  {2009})}\BibitemShut {NoStop}%
\bibitem [{\citenamefont {Kelleher}\ \emph {et~al.}(2017)\citenamefont
  {Kelleher}, \citenamefont {Menges}, \citenamefont {DePalma}, \citenamefont
  {Denton}, \citenamefont {Johnson}, \citenamefont {Weddle}, \citenamefont
  {Hirshberg},\ and\ \citenamefont {Gerber}}]{kelleher2017trapping}%
  \BibitemOpen
  \bibfield  {author} {\bibinfo {author} {\bibfnamefont {P.~J.}\ \bibnamefont
  {Kelleher}}, \bibinfo {author} {\bibfnamefont {F.~S.}\ \bibnamefont
  {Menges}}, \bibinfo {author} {\bibfnamefont {J.~W.}\ \bibnamefont {DePalma}},
  \bibinfo {author} {\bibfnamefont {J.~K.}\ \bibnamefont {Denton}}, \bibinfo
  {author} {\bibfnamefont {M.~A.}\ \bibnamefont {Johnson}}, \bibinfo {author}
  {\bibfnamefont {G.~H.}\ \bibnamefont {Weddle}}, \bibinfo {author}
  {\bibfnamefont {B.}~\bibnamefont {Hirshberg}},\ and\ \bibinfo {author}
  {\bibfnamefont {R.~B.}\ \bibnamefont {Gerber}},\ }\bibfield  {title}
  {\enquote {\bibinfo {title} {Trapping and structural characterization of the
  xno2{\textperiodcentered} no3--(x= cl, br, i) exit channel complexes in the
  water-mediated x--+ n2o5 reactions with cryogenic vibrational
  spectroscopy},}\ }\href@noop {} {\bibfield  {journal} {\bibinfo  {journal}
  {The Journal of Physical Chemistry Letters}\ }\textbf {\bibinfo {volume}
  {8}},\ \bibinfo {pages} {4710--4715} (\bibinfo {year} {2017})}\BibitemShut
  {NoStop}%
\bibitem [{\citenamefont {Karimova}\ \emph {et~al.}(2019)\citenamefont
  {Karimova}, \citenamefont {Chen}, \citenamefont {Gord}, \citenamefont
  {Staudt}, \citenamefont {Bertram}, \citenamefont {Nathanson},\ and\
  \citenamefont {Gerber}}]{karimova2019sn2}%
  \BibitemOpen
  \bibfield  {author} {\bibinfo {author} {\bibfnamefont {N.~V.}\ \bibnamefont
  {Karimova}}, \bibinfo {author} {\bibfnamefont {J.}~\bibnamefont {Chen}},
  \bibinfo {author} {\bibfnamefont {J.~R.}\ \bibnamefont {Gord}}, \bibinfo
  {author} {\bibfnamefont {S.}~\bibnamefont {Staudt}}, \bibinfo {author}
  {\bibfnamefont {T.~H.}\ \bibnamefont {Bertram}}, \bibinfo {author}
  {\bibfnamefont {G.~M.}\ \bibnamefont {Nathanson}},\ and\ \bibinfo {author}
  {\bibfnamefont {R.~B.}\ \bibnamefont {Gerber}},\ }\bibfield  {title}
  {\enquote {\bibinfo {title} {Sn2 reactions of n2o5 with ions in water:
  Microscopic mechanisms, intermediates, and products},}\ }\href@noop {}
  {\bibfield  {journal} {\bibinfo  {journal} {The Journal of Physical Chemistry
  A}\ }\textbf {\bibinfo {volume} {124}},\ \bibinfo {pages} {711--720}
  (\bibinfo {year} {2019})}\BibitemShut {NoStop}%
\bibitem [{\citenamefont {McCaslin}\ \emph {et~al.}(2023)\citenamefont
  {McCaslin}, \citenamefont {G{\"o}tz}, \citenamefont {Johnson},\ and\
  \citenamefont {Gerber}}]{mccaslin2023effects}%
  \BibitemOpen
  \bibfield  {author} {\bibinfo {author} {\bibfnamefont {L.~M.}\ \bibnamefont
  {McCaslin}}, \bibinfo {author} {\bibfnamefont {A.~W.}\ \bibnamefont
  {G{\"o}tz}}, \bibinfo {author} {\bibfnamefont {M.~A.}\ \bibnamefont
  {Johnson}},\ and\ \bibinfo {author} {\bibfnamefont {R.~B.}\ \bibnamefont
  {Gerber}},\ }\bibfield  {title} {\enquote {\bibinfo {title} {Effects of
  microhydration on the mechanisms of hydrolysis and cl- substitution in
  reactions of n2o5 and seawater},}\ }\href@noop {} {\bibfield  {journal}
  {\bibinfo  {journal} {ChemPhysChem}\ }\textbf {\bibinfo {volume} {24}},\
  \bibinfo {pages} {e202200819} (\bibinfo {year} {2023})}\BibitemShut {NoStop}%
\bibitem [{\citenamefont {Warshel}(1991)}]{warshel1991computer}%
  \BibitemOpen
  \bibfield  {author} {\bibinfo {author} {\bibfnamefont {A.}~\bibnamefont
  {Warshel}},\ }\bibfield  {title} {\enquote {\bibinfo {title} {Computer
  modelling of chemical reactions in enzymes and solutions},}\ }\href@noop {}
  {\  (\bibinfo {year} {1991})}\BibitemShut {NoStop}%
\bibitem [{\citenamefont {Hummer}(2004)}]{hummer2004transition}%
  \BibitemOpen
  \bibfield  {author} {\bibinfo {author} {\bibfnamefont {G.}~\bibnamefont
  {Hummer}},\ }\bibfield  {title} {\enquote {\bibinfo {title} {From transition
  paths to transition states and rate coefficients},}\ }\href@noop {}
  {\bibfield  {journal} {\bibinfo  {journal} {The Journal of chemical physics}\
  }\textbf {\bibinfo {volume} {120}},\ \bibinfo {pages} {516--523} (\bibinfo
  {year} {2004})}\BibitemShut {NoStop}%
\bibitem [{\citenamefont {Davidovits}\ \emph {et~al.}(2006)\citenamefont
  {Davidovits}, \citenamefont {Kolb}, \citenamefont {Williams}, \citenamefont
  {Jayne},\ and\ \citenamefont {Worsnop}}]{davidovits2006mass}%
  \BibitemOpen
  \bibfield  {author} {\bibinfo {author} {\bibfnamefont {P.}~\bibnamefont
  {Davidovits}}, \bibinfo {author} {\bibfnamefont {C.~E.}\ \bibnamefont
  {Kolb}}, \bibinfo {author} {\bibfnamefont {L.~R.}\ \bibnamefont {Williams}},
  \bibinfo {author} {\bibfnamefont {J.~T.}\ \bibnamefont {Jayne}},\ and\
  \bibinfo {author} {\bibfnamefont {D.~R.}\ \bibnamefont {Worsnop}},\
  }\bibfield  {title} {\enquote {\bibinfo {title} {Mass accommodation and
  chemical reactions at gas- liquid interfaces},}\ }\href@noop {} {\bibfield
  {journal} {\bibinfo  {journal} {Chemical reviews}\ }\textbf {\bibinfo
  {volume} {106}},\ \bibinfo {pages} {1323--1354} (\bibinfo {year}
  {2006})}\BibitemShut {NoStop}%
\bibitem [{\citenamefont {Danckwerts}(1951)}]{danckwerts1951absorption}%
  \BibitemOpen
  \bibfield  {author} {\bibinfo {author} {\bibfnamefont {P.}~\bibnamefont
  {Danckwerts}},\ }\bibfield  {title} {\enquote {\bibinfo {title} {Absorption
  by simultaneous diffusion and chemical reaction into particles of various
  shapes and into falling drops},}\ }\href@noop {} {\bibfield  {journal}
  {\bibinfo  {journal} {Transactions of the faraday society}\ }\textbf
  {\bibinfo {volume} {47}},\ \bibinfo {pages} {1014--1023} (\bibinfo {year}
  {1951})}\BibitemShut {NoStop}%
\bibitem [{\citenamefont {Ballard}\ and\ \citenamefont
  {Dellago}(2012)}]{ballard2012toward}%
  \BibitemOpen
  \bibfield  {author} {\bibinfo {author} {\bibfnamefont {A.~J.}\ \bibnamefont
  {Ballard}}\ and\ \bibinfo {author} {\bibfnamefont {C.}~\bibnamefont
  {Dellago}},\ }\bibfield  {title} {\enquote {\bibinfo {title} {Toward the
  mechanism of ionic dissociation in water},}\ }\href@noop {} {\bibfield
  {journal} {\bibinfo  {journal} {The Journal of Physical Chemistry B}\
  }\textbf {\bibinfo {volume} {116}},\ \bibinfo {pages} {13490--13497}
  (\bibinfo {year} {2012})}\BibitemShut {NoStop}%
\bibitem [{\citenamefont {Kattirtzi}, \citenamefont {Limmer},\ and\
  \citenamefont {Willard}(2017)}]{kattirtzi2017microscopic}%
  \BibitemOpen
  \bibfield  {author} {\bibinfo {author} {\bibfnamefont {J.~A.}\ \bibnamefont
  {Kattirtzi}}, \bibinfo {author} {\bibfnamefont {D.~T.}\ \bibnamefont
  {Limmer}},\ and\ \bibinfo {author} {\bibfnamefont {A.~P.}\ \bibnamefont
  {Willard}},\ }\bibfield  {title} {\enquote {\bibinfo {title} {Microscopic
  dynamics of charge separation at the aqueous electrochemical interface},}\
  }\href@noop {} {\bibfield  {journal} {\bibinfo  {journal} {Proceedings of the
  National Academy of Sciences}\ }\textbf {\bibinfo {volume} {114}},\ \bibinfo
  {pages} {13374--13379} (\bibinfo {year} {2017})}\BibitemShut {NoStop}%
\bibitem [{\citenamefont {Geissler}\ \emph {et~al.}(2001)\citenamefont
  {Geissler}, \citenamefont {Dellago}, \citenamefont {Chandler}, \citenamefont
  {Hutter},\ and\ \citenamefont {Parrinello}}]{geissler2001autoionization}%
  \BibitemOpen
  \bibfield  {author} {\bibinfo {author} {\bibfnamefont {P.~L.}\ \bibnamefont
  {Geissler}}, \bibinfo {author} {\bibfnamefont {C.}~\bibnamefont {Dellago}},
  \bibinfo {author} {\bibfnamefont {D.}~\bibnamefont {Chandler}}, \bibinfo
  {author} {\bibfnamefont {J.}~\bibnamefont {Hutter}},\ and\ \bibinfo {author}
  {\bibfnamefont {M.}~\bibnamefont {Parrinello}},\ }\bibfield  {title}
  {\enquote {\bibinfo {title} {Autoionization in liquid water},}\ }\href@noop
  {} {\bibfield  {journal} {\bibinfo  {journal} {Science}\ }\textbf {\bibinfo
  {volume} {291}},\ \bibinfo {pages} {2121--2124} (\bibinfo {year}
  {2001})}\BibitemShut {NoStop}%
\bibitem [{\citenamefont {Reischl}, \citenamefont {K{\"o}finger},\ and\
  \citenamefont {Dellago}(2009)}]{reischl2009statistics}%
  \BibitemOpen
  \bibfield  {author} {\bibinfo {author} {\bibfnamefont {B.}~\bibnamefont
  {Reischl}}, \bibinfo {author} {\bibfnamefont {J.}~\bibnamefont
  {K{\"o}finger}},\ and\ \bibinfo {author} {\bibfnamefont {C.}~\bibnamefont
  {Dellago}},\ }\bibfield  {title} {\enquote {\bibinfo {title} {The statistics
  of electric field fluctuations in liquid water},}\ }\href@noop {} {\bibfield
  {journal} {\bibinfo  {journal} {Molecular Physics}\ }\textbf {\bibinfo
  {volume} {107}},\ \bibinfo {pages} {495--502} (\bibinfo {year}
  {2009})}\BibitemShut {NoStop}%
\bibitem [{\citenamefont {Wang}\ \emph {et~al.}(2004)\citenamefont {Wang},
  \citenamefont {Wolf}, \citenamefont {Caldwell}, \citenamefont {Kollman},\
  and\ \citenamefont {Case}}]{wang2004development}%
  \BibitemOpen
  \bibfield  {author} {\bibinfo {author} {\bibfnamefont {J.}~\bibnamefont
  {Wang}}, \bibinfo {author} {\bibfnamefont {R.~M.}\ \bibnamefont {Wolf}},
  \bibinfo {author} {\bibfnamefont {J.~W.}\ \bibnamefont {Caldwell}}, \bibinfo
  {author} {\bibfnamefont {P.~A.}\ \bibnamefont {Kollman}},\ and\ \bibinfo
  {author} {\bibfnamefont {D.~A.}\ \bibnamefont {Case}},\ }\bibfield  {title}
  {\enquote {\bibinfo {title} {Development and testing of a general amber force
  field},}\ }\href@noop {} {\bibfield  {journal} {\bibinfo  {journal} {Journal
  of computational chemistry}\ }\textbf {\bibinfo {volume} {25}},\ \bibinfo
  {pages} {1157--1174} (\bibinfo {year} {2004})}\BibitemShut {NoStop}%
\bibitem [{\citenamefont {Ou}\ \emph {et~al.}(2013)\citenamefont {Ou},
  \citenamefont {Hu}, \citenamefont {Patel},\ and\ \citenamefont
  {Wan}}]{ou2013spherical}%
  \BibitemOpen
  \bibfield  {author} {\bibinfo {author} {\bibfnamefont {S.}~\bibnamefont
  {Ou}}, \bibinfo {author} {\bibfnamefont {Y.}~\bibnamefont {Hu}}, \bibinfo
  {author} {\bibfnamefont {S.}~\bibnamefont {Patel}},\ and\ \bibinfo {author}
  {\bibfnamefont {H.}~\bibnamefont {Wan}},\ }\bibfield  {title} {\enquote
  {\bibinfo {title} {Spherical monovalent ions at aqueous liquid--vapor
  interfaces: Interfacial stability and induced interface fluctuations},}\
  }\href@noop {} {\bibfield  {journal} {\bibinfo  {journal} {The Journal of
  Physical Chemistry B}\ }\textbf {\bibinfo {volume} {117}},\ \bibinfo {pages}
  {11732--11742} (\bibinfo {year} {2013})}\BibitemShut {NoStop}%
\bibitem [{\citenamefont {Schlegel}\ and\ \citenamefont
  {Sonnenberg}(2006)}]{schlegel2006empirical}%
  \BibitemOpen
  \bibfield  {author} {\bibinfo {author} {\bibfnamefont {H.~B.}\ \bibnamefont
  {Schlegel}}\ and\ \bibinfo {author} {\bibfnamefont {J.~L.}\ \bibnamefont
  {Sonnenberg}},\ }\bibfield  {title} {\enquote {\bibinfo {title} {Empirical
  valence-bond models for reactive potential energy surfaces using distributed
  gaussians},}\ }\href@noop {} {\bibfield  {journal} {\bibinfo  {journal}
  {Journal of chemical theory and computation}\ }\textbf {\bibinfo {volume}
  {2}},\ \bibinfo {pages} {905--911} (\bibinfo {year} {2006})}\BibitemShut
  {NoStop}%
\bibitem [{\citenamefont {Mardirossian}\ and\ \citenamefont
  {Head-Gordon}(2014)}]{mardirossian2014omegab97x}%
  \BibitemOpen
  \bibfield  {author} {\bibinfo {author} {\bibfnamefont {N.}~\bibnamefont
  {Mardirossian}}\ and\ \bibinfo {author} {\bibfnamefont {M.}~\bibnamefont
  {Head-Gordon}},\ }\bibfield  {title} {\enquote {\bibinfo {title}
  {$\omega$b97x-v: A 10-parameter, range-separated hybrid, generalized gradient
  approximation density functional with nonlocal correlation, designed by a
  survival-of-the-fittest strategy},}\ }\href@noop {} {\bibfield  {journal}
  {\bibinfo  {journal} {Physical Chemistry Chemical Physics}\ }\textbf
  {\bibinfo {volume} {16}},\ \bibinfo {pages} {9904--9924} (\bibinfo {year}
  {2014})}\BibitemShut {NoStop}%
\bibitem [{\citenamefont {Wu}, \citenamefont {Tepper},\ and\ \citenamefont
  {Voth}(2006)}]{wu2006flexible}%
  \BibitemOpen
  \bibfield  {author} {\bibinfo {author} {\bibfnamefont {Y.}~\bibnamefont
  {Wu}}, \bibinfo {author} {\bibfnamefont {H.~L.}\ \bibnamefont {Tepper}},\
  and\ \bibinfo {author} {\bibfnamefont {G.~A.}\ \bibnamefont {Voth}},\
  }\bibfield  {title} {\enquote {\bibinfo {title} {Flexible simple point-charge
  water model with improved liquid-state properties},}\ }\href@noop {}
  {\bibfield  {journal} {\bibinfo  {journal} {The Journal of chemical physics}\
  }\textbf {\bibinfo {volume} {124}} (\bibinfo {year} {2006})}\BibitemShut
  {NoStop}%
\bibitem [{\citenamefont {Frenkel}\ and\ \citenamefont
  {Smit}(2023)}]{frenkel2023understanding}%
  \BibitemOpen
  \bibfield  {author} {\bibinfo {author} {\bibfnamefont {D.}~\bibnamefont
  {Frenkel}}\ and\ \bibinfo {author} {\bibfnamefont {B.}~\bibnamefont {Smit}},\
  }\href@noop {} {\emph {\bibinfo {title} {Understanding molecular simulation:
  from algorithms to applications}}}\ (\bibinfo  {publisher} {Elsevier},\
  \bibinfo {year} {2023})\BibitemShut {NoStop}%
\bibitem [{\citenamefont {Kumar}\ \emph {et~al.}(1995)\citenamefont {Kumar},
  \citenamefont {Rosenberg}, \citenamefont {Bouzida}, \citenamefont
  {Swendsen},\ and\ \citenamefont {Kollman}}]{kumar1995multidimensional}%
  \BibitemOpen
  \bibfield  {author} {\bibinfo {author} {\bibfnamefont {S.}~\bibnamefont
  {Kumar}}, \bibinfo {author} {\bibfnamefont {J.~M.}\ \bibnamefont
  {Rosenberg}}, \bibinfo {author} {\bibfnamefont {D.}~\bibnamefont {Bouzida}},
  \bibinfo {author} {\bibfnamefont {R.~H.}\ \bibnamefont {Swendsen}},\ and\
  \bibinfo {author} {\bibfnamefont {P.~A.}\ \bibnamefont {Kollman}},\
  }\bibfield  {title} {\enquote {\bibinfo {title} {Multidimensional free-energy
  calculations using the weighted histogram analysis method},}\ }\href@noop {}
  {\bibfield  {journal} {\bibinfo  {journal} {Journal of Computational
  Chemistry}\ }\textbf {\bibinfo {volume} {16}},\ \bibinfo {pages} {1339--1350}
  (\bibinfo {year} {1995})}\BibitemShut {NoStop}%
\bibitem [{\citenamefont {Loche}, \citenamefont {Bonthuis},\ and\ \citenamefont
  {Netz}(2022)}]{loche2022molecular}%
  \BibitemOpen
  \bibfield  {author} {\bibinfo {author} {\bibfnamefont {P.}~\bibnamefont
  {Loche}}, \bibinfo {author} {\bibfnamefont {D.~J.}\ \bibnamefont
  {Bonthuis}},\ and\ \bibinfo {author} {\bibfnamefont {R.~R.}\ \bibnamefont
  {Netz}},\ }\bibfield  {title} {\enquote {\bibinfo {title} {Molecular dynamics
  simulations of the evaporation of hydrated ions from aqueous solution},}\
  }\href@noop {} {\bibfield  {journal} {\bibinfo  {journal} {Communications
  Chemistry}\ }\textbf {\bibinfo {volume} {5}},\ \bibinfo {pages} {55}
  (\bibinfo {year} {2022})}\BibitemShut {NoStop}%
\bibitem [{\citenamefont {Fried}\ \emph {et~al.}(1994)\citenamefont {Fried},
  \citenamefont {Henry}, \citenamefont {Calvert},\ and\ \citenamefont
  {Mozurkewich}}]{fried1994reaction}%
  \BibitemOpen
  \bibfield  {author} {\bibinfo {author} {\bibfnamefont {A.}~\bibnamefont
  {Fried}}, \bibinfo {author} {\bibfnamefont {B.~E.}\ \bibnamefont {Henry}},
  \bibinfo {author} {\bibfnamefont {J.~G.}\ \bibnamefont {Calvert}},\ and\
  \bibinfo {author} {\bibfnamefont {M.}~\bibnamefont {Mozurkewich}},\
  }\bibfield  {title} {\enquote {\bibinfo {title} {The reaction probability of
  n2o5 with sulfuric acid aerosols at stratospheric temperatures and
  compositions},}\ }\href@noop {} {\bibfield  {journal} {\bibinfo  {journal}
  {Journal of Geophysical Research: Atmospheres}\ }\textbf {\bibinfo {volume}
  {99}},\ \bibinfo {pages} {3517--3532} (\bibinfo {year} {1994})}\BibitemShut
  {NoStop}%
\bibitem [{\citenamefont {Behnke}\ \emph {et~al.}(1997)\citenamefont {Behnke},
  \citenamefont {George}, \citenamefont {Scheer},\ and\ \citenamefont
  {Zetzsch}}]{behnke1997production}%
  \BibitemOpen
  \bibfield  {author} {\bibinfo {author} {\bibfnamefont {W.}~\bibnamefont
  {Behnke}}, \bibinfo {author} {\bibfnamefont {C.}~\bibnamefont {George}},
  \bibinfo {author} {\bibfnamefont {V.}~\bibnamefont {Scheer}},\ and\ \bibinfo
  {author} {\bibfnamefont {C.}~\bibnamefont {Zetzsch}},\ }\bibfield  {title}
  {\enquote {\bibinfo {title} {Production and decay of clno2 from the reaction
  of gaseous n2o5 with nacl solution: Bulk and aerosol experiments},}\
  }\href@noop {} {\bibfield  {journal} {\bibinfo  {journal} {Journal of
  Geophysical Research: Atmospheres}\ }\textbf {\bibinfo {volume} {102}},\
  \bibinfo {pages} {3795--3804} (\bibinfo {year} {1997})}\BibitemShut {NoStop}%
\bibitem [{\citenamefont {Hamlin}, \citenamefont {Swart},\ and\ \citenamefont
  {Bickelhaupt}(2018)}]{hamlin2018nucleophilic}%
  \BibitemOpen
  \bibfield  {author} {\bibinfo {author} {\bibfnamefont {T.~A.}\ \bibnamefont
  {Hamlin}}, \bibinfo {author} {\bibfnamefont {M.}~\bibnamefont {Swart}},\ and\
  \bibinfo {author} {\bibfnamefont {F.~M.}\ \bibnamefont {Bickelhaupt}},\
  }\bibfield  {title} {\enquote {\bibinfo {title} {Nucleophilic substitution
  (sn2): dependence on nucleophile, leaving group, central atom, substituents,
  and solvent},}\ }\href@noop {} {\bibfield  {journal} {\bibinfo  {journal}
  {ChemPhysChem}\ }\textbf {\bibinfo {volume} {19}},\ \bibinfo {pages}
  {1315--1330} (\bibinfo {year} {2018})}\BibitemShut {NoStop}%
\bibitem [{\citenamefont {Benjamin}(2008)}]{benjamin2008empirical}%
  \BibitemOpen
  \bibfield  {author} {\bibinfo {author} {\bibfnamefont {I.}~\bibnamefont
  {Benjamin}},\ }\bibfield  {title} {\enquote {\bibinfo {title} {Empirical
  valence bond model of an sn2 reaction in polar and nonpolar solvents},}\
  }\href@noop {} {\bibfield  {journal} {\bibinfo  {journal} {The Journal of
  chemical physics}\ }\textbf {\bibinfo {volume} {129}} (\bibinfo {year}
  {2008})}\BibitemShut {NoStop}%
\bibitem [{\citenamefont {Shaffer}(2013)}]{shaffer2013ion}%
  \BibitemOpen
  \bibfield  {author} {\bibinfo {author} {\bibfnamefont {P.~R.}\ \bibnamefont
  {Shaffer}},\ }\href@noop {} {\emph {\bibinfo {title} {Ion solvation at
  air-water interfaces}}}\ (\bibinfo  {publisher} {University of California,
  Berkeley},\ \bibinfo {year} {2013})\BibitemShut {NoStop}%
\bibitem [{\citenamefont {Jungwirth}\ and\ \citenamefont
  {Tobias}(2001)}]{jungwirth2001molecular}%
  \BibitemOpen
  \bibfield  {author} {\bibinfo {author} {\bibfnamefont {P.}~\bibnamefont
  {Jungwirth}}\ and\ \bibinfo {author} {\bibfnamefont {D.~J.}\ \bibnamefont
  {Tobias}},\ }\bibfield  {title} {\enquote {\bibinfo {title} {Molecular
  structure of salt solutions: A new view of the interface with implications
  for heterogeneous atmospheric chemistry},}\ }\href@noop {} {\bibfield
  {journal} {\bibinfo  {journal} {The Journal of Physical Chemistry B}\
  }\textbf {\bibinfo {volume} {105}},\ \bibinfo {pages} {10468--10472}
  (\bibinfo {year} {2001})}\BibitemShut {NoStop}%
\bibitem [{\citenamefont {Otten}\ \emph {et~al.}(2012)\citenamefont {Otten},
  \citenamefont {Shaffer}, \citenamefont {Geissler},\ and\ \citenamefont
  {Saykally}}]{otten2012elucidating}%
  \BibitemOpen
  \bibfield  {author} {\bibinfo {author} {\bibfnamefont {D.~E.}\ \bibnamefont
  {Otten}}, \bibinfo {author} {\bibfnamefont {P.~R.}\ \bibnamefont {Shaffer}},
  \bibinfo {author} {\bibfnamefont {P.~L.}\ \bibnamefont {Geissler}},\ and\
  \bibinfo {author} {\bibfnamefont {R.~J.}\ \bibnamefont {Saykally}},\
  }\bibfield  {title} {\enquote {\bibinfo {title} {Elucidating the mechanism of
  selective ion adsorption to the liquid water surface},}\ }\href@noop {}
  {\bibfield  {journal} {\bibinfo  {journal} {Proceedings of the National
  Academy of Sciences}\ }\textbf {\bibinfo {volume} {109}},\ \bibinfo {pages}
  {701--705} (\bibinfo {year} {2012})}\BibitemShut {NoStop}%
\bibitem [{\citenamefont {Schile}\ and\ \citenamefont
  {Limmer}(2019)}]{schile2019rate}%
  \BibitemOpen
  \bibfield  {author} {\bibinfo {author} {\bibfnamefont {A.~J.}\ \bibnamefont
  {Schile}}\ and\ \bibinfo {author} {\bibfnamefont {D.~T.}\ \bibnamefont
  {Limmer}},\ }\bibfield  {title} {\enquote {\bibinfo {title} {Rate constants
  in spatially inhomogeneous systems},}\ }\href@noop {} {\bibfield  {journal}
  {\bibinfo  {journal} {The Journal of chemical physics}\ }\textbf {\bibinfo
  {volume} {150}} (\bibinfo {year} {2019})}\BibitemShut {NoStop}%
\bibitem [{\citenamefont {Laidler}\ and\ \citenamefont
  {King}(1983)}]{laidler1983development}%
  \BibitemOpen
  \bibfield  {author} {\bibinfo {author} {\bibfnamefont {K.~J.}\ \bibnamefont
  {Laidler}}\ and\ \bibinfo {author} {\bibfnamefont {M.~C.}\ \bibnamefont
  {King}},\ }\bibfield  {title} {\enquote {\bibinfo {title} {The development of
  transition-state theory},}\ }\href@noop {} {\bibfield  {journal} {\bibinfo
  {journal} {J. phys. Chem}\ }\textbf {\bibinfo {volume} {87}},\ \bibinfo
  {pages} {2657--2664} (\bibinfo {year} {1983})}\BibitemShut {NoStop}%
\bibitem [{\citenamefont {Berne}, \citenamefont {Borkovec},\ and\ \citenamefont
  {Straub}(1988)}]{berne1988classical}%
  \BibitemOpen
  \bibfield  {author} {\bibinfo {author} {\bibfnamefont {B.~J.}\ \bibnamefont
  {Berne}}, \bibinfo {author} {\bibfnamefont {M.}~\bibnamefont {Borkovec}},\
  and\ \bibinfo {author} {\bibfnamefont {J.~E.}\ \bibnamefont {Straub}},\
  }\bibfield  {title} {\enquote {\bibinfo {title} {Classical and modern methods
  in reaction rate theory},}\ }\href@noop {} {\bibfield  {journal} {\bibinfo
  {journal} {The Journal of Physical Chemistry}\ }\textbf {\bibinfo {volume}
  {92}},\ \bibinfo {pages} {3711--3725} (\bibinfo {year} {1988})}\BibitemShut
  {NoStop}%
\bibitem [{\citenamefont {Chandler}(1978)}]{chandler1978statistical}%
  \BibitemOpen
  \bibfield  {author} {\bibinfo {author} {\bibfnamefont {D.}~\bibnamefont
  {Chandler}},\ }\bibfield  {title} {\enquote {\bibinfo {title} {Statistical
  mechanics of isomerization dynamics in liquids and the transition state
  approximation},}\ }\href@noop {} {\bibfield  {journal} {\bibinfo  {journal}
  {The Journal of Chemical Physics}\ }\textbf {\bibinfo {volume} {68}},\
  \bibinfo {pages} {2959--2970} (\bibinfo {year} {1978})}\BibitemShut {NoStop}%
\bibitem [{\citenamefont {Danckwerts}\ \emph {et~al.}(1970)\citenamefont
  {Danckwerts} \emph {et~al.}}]{danckwerts1970gas}%
  \BibitemOpen
  \bibfield  {author} {\bibinfo {author} {\bibfnamefont {P.~V.}\ \bibnamefont
  {Danckwerts}} \emph {et~al.},\ }\bibfield  {title} {\enquote {\bibinfo
  {title} {Gas-liquid reactions},}\ }\href@noop {} {\  (\bibinfo {year}
  {1970})}\BibitemShut {NoStop}%
\bibitem [{\citenamefont {Worsnop}\ \emph {et~al.}(1989)\citenamefont
  {Worsnop}, \citenamefont {Zahniser}, \citenamefont {Kolb}, \citenamefont
  {Gardner}, \citenamefont {Watson}, \citenamefont {Van~Doren}, \citenamefont
  {Jayne},\ and\ \citenamefont {Davidovits}}]{worsnop1989temperature}%
  \BibitemOpen
  \bibfield  {author} {\bibinfo {author} {\bibfnamefont {D.~R.}\ \bibnamefont
  {Worsnop}}, \bibinfo {author} {\bibfnamefont {M.~S.}\ \bibnamefont
  {Zahniser}}, \bibinfo {author} {\bibfnamefont {C.~E.}\ \bibnamefont {Kolb}},
  \bibinfo {author} {\bibfnamefont {J.~A.}\ \bibnamefont {Gardner}}, \bibinfo
  {author} {\bibfnamefont {L.~R.}\ \bibnamefont {Watson}}, \bibinfo {author}
  {\bibfnamefont {J.~M.}\ \bibnamefont {Van~Doren}}, \bibinfo {author}
  {\bibfnamefont {J.~T.}\ \bibnamefont {Jayne}},\ and\ \bibinfo {author}
  {\bibfnamefont {P.}~\bibnamefont {Davidovits}},\ }\bibfield  {title}
  {\enquote {\bibinfo {title} {The temperature dependence of mass accommodation
  of sulfur dioxide and hydrogen peroxide on aqueous surfaces},}\ }\href@noop
  {} {\bibfield  {journal} {\bibinfo  {journal} {The Journal of Physical
  Chemistry}\ }\textbf {\bibinfo {volume} {93}},\ \bibinfo {pages} {1159--1172}
  (\bibinfo {year} {1989})}\BibitemShut {NoStop}%
\bibitem [{\citenamefont {Polley}, \citenamefont {Wilson},\ and\ \citenamefont
  {Limmer}(2024)}]{polley2024statistical}%
  \BibitemOpen
  \bibfield  {author} {\bibinfo {author} {\bibfnamefont {K.}~\bibnamefont
  {Polley}}, \bibinfo {author} {\bibfnamefont {K.~R.}\ \bibnamefont {Wilson}},\
  and\ \bibinfo {author} {\bibfnamefont {D.~T.}\ \bibnamefont {Limmer}},\
  }\bibfield  {title} {\enquote {\bibinfo {title} {On the statistical mechanics
  of mass accommodation at liquid-vapor interfaces},}\ }\href@noop {}
  {\bibfield  {journal} {\bibinfo  {journal} {arXiv:2401.13234}\ } (\bibinfo
  {year} {2024})}\BibitemShut {NoStop}%
\bibitem [{\citenamefont {Kregel}\ \emph {et~al.}(2023)\citenamefont {Kregel},
  \citenamefont {Derrah}, \citenamefont {Moon}, \citenamefont {Limmer},
  \citenamefont {Nathanson},\ and\ \citenamefont {Bertram}}]{kregel2023weak}%
  \BibitemOpen
  \bibfield  {author} {\bibinfo {author} {\bibfnamefont {S.~J.}\ \bibnamefont
  {Kregel}}, \bibinfo {author} {\bibfnamefont {T.~F.}\ \bibnamefont {Derrah}},
  \bibinfo {author} {\bibfnamefont {S.}~\bibnamefont {Moon}}, \bibinfo {author}
  {\bibfnamefont {D.~T.}\ \bibnamefont {Limmer}}, \bibinfo {author}
  {\bibfnamefont {G.~M.}\ \bibnamefont {Nathanson}},\ and\ \bibinfo {author}
  {\bibfnamefont {T.~H.}\ \bibnamefont {Bertram}},\ }\bibfield  {title}
  {\enquote {\bibinfo {title} {Weak temperature dependence of the relative
  rates of chlorination and hydrolysis of n2o5 in nacl--water solutions},}\
  }\href@noop {} {\bibfield  {journal} {\bibinfo  {journal} {The Journal of
  Physical Chemistry A}\ }\textbf {\bibinfo {volume} {127}},\ \bibinfo {pages}
  {1675--1685} (\bibinfo {year} {2023})}\BibitemShut {NoStop}%
\bibitem [{\citenamefont {Qiu}\ \emph {et~al.}(2021)\citenamefont {Qiu},
  \citenamefont {Wei}, \citenamefont {Nie},\ and\ \citenamefont
  {Cooks}}]{qiu2021reaction}%
  \BibitemOpen
  \bibfield  {author} {\bibinfo {author} {\bibfnamefont {L.}~\bibnamefont
  {Qiu}}, \bibinfo {author} {\bibfnamefont {Z.}~\bibnamefont {Wei}}, \bibinfo
  {author} {\bibfnamefont {H.}~\bibnamefont {Nie}},\ and\ \bibinfo {author}
  {\bibfnamefont {R.~G.}\ \bibnamefont {Cooks}},\ }\bibfield  {title} {\enquote
  {\bibinfo {title} {Reaction acceleration promoted by partial solvation at the
  gas/solution interface},}\ }\href@noop {} {\bibfield  {journal} {\bibinfo
  {journal} {ChemPlusChem}\ }\textbf {\bibinfo {volume} {86}},\ \bibinfo
  {pages} {1362--1365} (\bibinfo {year} {2021})}\BibitemShut {NoStop}%
\bibitem [{\citenamefont {Weigend}\ and\ \citenamefont
  {Ahlrichs}(2005)}]{weigend2005balanced}%
  \BibitemOpen
  \bibfield  {author} {\bibinfo {author} {\bibfnamefont {F.}~\bibnamefont
  {Weigend}}\ and\ \bibinfo {author} {\bibfnamefont {R.}~\bibnamefont
  {Ahlrichs}},\ }\bibfield  {title} {\enquote {\bibinfo {title} {Balanced basis
  sets of split valence, triple zeta valence and quadruple zeta valence quality
  for h to rn: Design and assessment of accuracy},}\ }\href@noop {} {\bibfield
  {journal} {\bibinfo  {journal} {Physical Chemistry Chemical Physics}\
  }\textbf {\bibinfo {volume} {7}},\ \bibinfo {pages} {3297--3305} (\bibinfo
  {year} {2005})}\BibitemShut {NoStop}%
\bibitem [{\citenamefont {Behn}\ \emph {et~al.}(2011)\citenamefont {Behn},
  \citenamefont {Zimmerman}, \citenamefont {Bell},\ and\ \citenamefont
  {Head-Gordon}}]{behn2011efficient}%
  \BibitemOpen
  \bibfield  {author} {\bibinfo {author} {\bibfnamefont {A.}~\bibnamefont
  {Behn}}, \bibinfo {author} {\bibfnamefont {P.~M.}\ \bibnamefont {Zimmerman}},
  \bibinfo {author} {\bibfnamefont {A.~T.}\ \bibnamefont {Bell}},\ and\
  \bibinfo {author} {\bibfnamefont {M.}~\bibnamefont {Head-Gordon}},\
  }\bibfield  {title} {\enquote {\bibinfo {title} {Efficient exploration of
  reaction paths via a freezing string method},}\ }\href@noop {} {\bibfield
  {journal} {\bibinfo  {journal} {The Journal of chemical physics}\ }\textbf
  {\bibinfo {volume} {135}} (\bibinfo {year} {2011})}\BibitemShut {NoStop}%
\bibitem [{\citenamefont {Fukui}(1970)}]{fukui1970formulation}%
  \BibitemOpen
  \bibfield  {author} {\bibinfo {author} {\bibfnamefont {K.}~\bibnamefont
  {Fukui}},\ }\bibfield  {title} {\enquote {\bibinfo {title} {Formulation of
  the reaction coordinate},}\ }\href@noop {} {\bibfield  {journal} {\bibinfo
  {journal} {The Journal of Physical Chemistry}\ }\textbf {\bibinfo {volume}
  {74}},\ \bibinfo {pages} {4161--4163} (\bibinfo {year} {1970})}\BibitemShut
  {NoStop}%
\bibitem [{\citenamefont {Shao}\ \emph {et~al.}(2015)\citenamefont {Shao},
  \citenamefont {Gan}, \citenamefont {Epifanovsky}, \citenamefont {Gilbert},
  \citenamefont {Wormit}, \citenamefont {Kussmann}, \citenamefont {Lange},
  \citenamefont {Behn}, \citenamefont {Deng}, \citenamefont {Feng} \emph
  {et~al.}}]{shao2015advances}%
  \BibitemOpen
  \bibfield  {author} {\bibinfo {author} {\bibfnamefont {Y.}~\bibnamefont
  {Shao}}, \bibinfo {author} {\bibfnamefont {Z.}~\bibnamefont {Gan}}, \bibinfo
  {author} {\bibfnamefont {E.}~\bibnamefont {Epifanovsky}}, \bibinfo {author}
  {\bibfnamefont {A.~T.}\ \bibnamefont {Gilbert}}, \bibinfo {author}
  {\bibfnamefont {M.}~\bibnamefont {Wormit}}, \bibinfo {author} {\bibfnamefont
  {J.}~\bibnamefont {Kussmann}}, \bibinfo {author} {\bibfnamefont {A.~W.}\
  \bibnamefont {Lange}}, \bibinfo {author} {\bibfnamefont {A.}~\bibnamefont
  {Behn}}, \bibinfo {author} {\bibfnamefont {J.}~\bibnamefont {Deng}}, \bibinfo
  {author} {\bibfnamefont {X.}~\bibnamefont {Feng}}, \emph {et~al.},\
  }\bibfield  {title} {\enquote {\bibinfo {title} {Advances in molecular
  quantum chemistry contained in the q-chem 4 program package},}\ }\href@noop
  {} {\bibfield  {journal} {\bibinfo  {journal} {Molecular Physics}\ }\textbf
  {\bibinfo {volume} {113}},\ \bibinfo {pages} {184--215} (\bibinfo {year}
  {2015})}\BibitemShut {NoStop}%
\bibitem [{\citenamefont {Shurki}\ \emph {et~al.}(2015)\citenamefont {Shurki},
  \citenamefont {Derat}, \citenamefont {Barrozo},\ and\ \citenamefont
  {Kamerlin}}]{shurki2015valence}%
  \BibitemOpen
  \bibfield  {author} {\bibinfo {author} {\bibfnamefont {A.}~\bibnamefont
  {Shurki}}, \bibinfo {author} {\bibfnamefont {E.}~\bibnamefont {Derat}},
  \bibinfo {author} {\bibfnamefont {A.}~\bibnamefont {Barrozo}},\ and\ \bibinfo
  {author} {\bibfnamefont {S.~C.~L.}\ \bibnamefont {Kamerlin}},\ }\bibfield
  {title} {\enquote {\bibinfo {title} {How valence bond theory can help you
  understand your (bio) chemical reaction},}\ }\href@noop {} {\bibfield
  {journal} {\bibinfo  {journal} {Chemical Society Reviews}\ }\textbf {\bibinfo
  {volume} {44}},\ \bibinfo {pages} {1037--1052} (\bibinfo {year}
  {2015})}\BibitemShut {NoStop}%
\bibitem [{\citenamefont {Warshel}\ and\ \citenamefont
  {Weiss}(1980)}]{warshel1980empirical}%
  \BibitemOpen
  \bibfield  {author} {\bibinfo {author} {\bibfnamefont {A.}~\bibnamefont
  {Warshel}}\ and\ \bibinfo {author} {\bibfnamefont {R.~M.}\ \bibnamefont
  {Weiss}},\ }\bibfield  {title} {\enquote {\bibinfo {title} {An empirical
  valence bond approach for comparing reactions in solutions and in enzymes},}\
  }\href@noop {} {\bibfield  {journal} {\bibinfo  {journal} {Journal of the
  American Chemical Society}\ }\textbf {\bibinfo {volume} {102}},\ \bibinfo
  {pages} {6218--6226} (\bibinfo {year} {1980})}\BibitemShut {NoStop}%
\bibitem [{\citenamefont {Cornell}\ \emph {et~al.}(2002)\citenamefont
  {Cornell}, \citenamefont {Cieplak}, \citenamefont {Bayly},\ and\
  \citenamefont {Kollman}}]{cornell2002application}%
  \BibitemOpen
  \bibfield  {author} {\bibinfo {author} {\bibfnamefont {W.~D.}\ \bibnamefont
  {Cornell}}, \bibinfo {author} {\bibfnamefont {P.}~\bibnamefont {Cieplak}},
  \bibinfo {author} {\bibfnamefont {C.~I.}\ \bibnamefont {Bayly}},\ and\
  \bibinfo {author} {\bibfnamefont {P.~A.}\ \bibnamefont {Kollman}},\
  }\bibfield  {title} {\enquote {\bibinfo {title} {Application of resp charges
  to calculate conformational energies, hydrogen bond energies, and free
  energies of solvation},}\ }\href@noop {} {\bibfield  {journal} {\bibinfo
  {journal} {Journal of the American Chemical Society}\ }\textbf {\bibinfo
  {volume} {115}},\ \bibinfo {pages} {9620--9631} (\bibinfo {year}
  {2002})}\BibitemShut {NoStop}%
\bibitem [{\citenamefont {Cossi}\ \emph {et~al.}(2003)\citenamefont {Cossi},
  \citenamefont {Rega}, \citenamefont {Scalmani},\ and\ \citenamefont
  {Barone}}]{cossi2003energies}%
  \BibitemOpen
  \bibfield  {author} {\bibinfo {author} {\bibfnamefont {M.}~\bibnamefont
  {Cossi}}, \bibinfo {author} {\bibfnamefont {N.}~\bibnamefont {Rega}},
  \bibinfo {author} {\bibfnamefont {G.}~\bibnamefont {Scalmani}},\ and\
  \bibinfo {author} {\bibfnamefont {V.}~\bibnamefont {Barone}},\ }\bibfield
  {title} {\enquote {\bibinfo {title} {Energies, structures, and electronic
  properties of molecules in solution with the c-pcm solvation model},}\
  }\href@noop {} {\bibfield  {journal} {\bibinfo  {journal} {Journal of
  computational chemistry}\ }\textbf {\bibinfo {volume} {24}},\ \bibinfo
  {pages} {669--681} (\bibinfo {year} {2003})}\BibitemShut {NoStop}%
\bibitem [{\citenamefont {Leontyev}\ and\ \citenamefont
  {Stuchebrukhov}(2011)}]{leontyev2011accounting}%
  \BibitemOpen
  \bibfield  {author} {\bibinfo {author} {\bibfnamefont {I.}~\bibnamefont
  {Leontyev}}\ and\ \bibinfo {author} {\bibfnamefont {A.}~\bibnamefont
  {Stuchebrukhov}},\ }\bibfield  {title} {\enquote {\bibinfo {title}
  {Accounting for electronic polarization in non-polarizable force fields},}\
  }\href@noop {} {\bibfield  {journal} {\bibinfo  {journal} {Physical Chemistry
  Chemical Physics}\ }\textbf {\bibinfo {volume} {13}},\ \bibinfo {pages}
  {2613--2626} (\bibinfo {year} {2011})}\BibitemShut {NoStop}%
\bibitem [{\citenamefont {Thompson}\ \emph {et~al.}(2022)\citenamefont
  {Thompson}, \citenamefont {Aktulga}, \citenamefont {Berger}, \citenamefont
  {Bolintineanu}, \citenamefont {Brown}, \citenamefont {Crozier}, \citenamefont
  {in't Veld}, \citenamefont {Kohlmeyer}, \citenamefont {Moore}, \citenamefont
  {Nguyen} \emph {et~al.}}]{thompson2022lammps}%
  \BibitemOpen
  \bibfield  {author} {\bibinfo {author} {\bibfnamefont {A.~P.}\ \bibnamefont
  {Thompson}}, \bibinfo {author} {\bibfnamefont {H.~M.}\ \bibnamefont
  {Aktulga}}, \bibinfo {author} {\bibfnamefont {R.}~\bibnamefont {Berger}},
  \bibinfo {author} {\bibfnamefont {D.~S.}\ \bibnamefont {Bolintineanu}},
  \bibinfo {author} {\bibfnamefont {W.~M.}\ \bibnamefont {Brown}}, \bibinfo
  {author} {\bibfnamefont {P.~S.}\ \bibnamefont {Crozier}}, \bibinfo {author}
  {\bibfnamefont {P.~J.}\ \bibnamefont {in't Veld}}, \bibinfo {author}
  {\bibfnamefont {A.}~\bibnamefont {Kohlmeyer}}, \bibinfo {author}
  {\bibfnamefont {S.~G.}\ \bibnamefont {Moore}}, \bibinfo {author}
  {\bibfnamefont {T.~D.}\ \bibnamefont {Nguyen}}, \emph {et~al.},\ }\bibfield
  {title} {\enquote {\bibinfo {title} {Lammps-a flexible simulation tool for
  particle-based materials modeling at the atomic, meso, and continuum
  scales},}\ }\href@noop {} {\bibfield  {journal} {\bibinfo  {journal}
  {Computer Physics Communications}\ }\textbf {\bibinfo {volume} {271}},\
  \bibinfo {pages} {108171} (\bibinfo {year} {2022})}\BibitemShut {NoStop}%
\bibitem [{\citenamefont {Grossfield}(2011)}]{grossfield2011wham}%
  \BibitemOpen
  \bibfield  {author} {\bibinfo {author} {\bibfnamefont {A.}~\bibnamefont
  {Grossfield}},\ }\bibfield  {title} {\enquote {\bibinfo {title} {Wham: an
  implementation of the weighted histogram analysis method},}\ }\href@noop {}
  {\bibfield  {journal} {\bibinfo  {journal} {University of Rochester, USA}\ }
  (\bibinfo {year} {2011})}\BibitemShut {NoStop}%
\end{thebibliography}%


\appendix

\section{\label{sec:modeling}Empirical Valance Bond Model for Chlorination Reaction}

The valence bond theory describes $\mathrm{S_N2}$-type charge transfer process as an adiabatic transition among low-lying valence-bond diabatic states\cite{shurki2015valence}. The empirical version of valence bond theory uses classical force-fields as an ansatz for elements of the quantum Hamiltonian matrix\cite{warshel1980empirical}. 
Here, two low-lying states corresponding to the reactant ($H_{11}$) and the product ($H_{22}$) state of the reaction (R2) are used to construct the adiabatic potential energy surface (PES), $H_{ad}$,
\begin{align}
    H_{ad} = \frac{1}{2} \left(H_{11} + H_{22}\right) - \frac{1}{2} \sqrt{(H_{11}-H_{22})^2 +4H_{12}^2} \, ,
\end{align}
where $H_{12}$ is the off-diagonal element or diabatic coupling between two states. The adiabatic PES is computed from $w\mathrm{B97X-V/DEF2-TZVPD}$ level of theory\cite{weigend2005balanced, mardirossian2014omegab97x}. All $ab-initio$ calculations are done with Qchem 5.4.2\cite{shao2015advances}.

Diabatic Hamiltonian elements are modeled by re-parameterizing the generalized Amber force field\cite{wang2004development} for the chemical species to reproduce the \textit{ab-initio} structures and the interfacial propensity around the air-water interface. Equilibrium bond distances and angles are modified to reproduce gas-phase DFT-optimized structures. Interactions of bond-forming ($\mathrm{Cl-N}$) and bond-breaking ($\mathrm{N-O}$) pairs are treated by replacing harmonic bonding interaction with a Morse potential fitted to the rigid-scan of bond length, and the non-bonded Lennard-Jones potential is changed to an exponential function to avoid an artificially strong short-ranged repulsive interaction. Dihedral interaction of $\mathrm{N_2O_5}$ is replaced to a sum of periodic functions to reproduce the rigid-scan of dihedral angles. Electrostatic interactions are modeled with point-charges centered at atomic centers, whose values are determined from the restricted electrostatic potential  calculation\cite{cornell2002application} with the conductor-like polarizable continuum  implicit solvent model\cite{cossi2003energies}. The size of the solute cavity, a union of atom-centered spheres, is used as a control parameter for the optimization. The reversible work profile evaluated from the MB-nrg model\cite{cruzeiro2022uptake} is the reference for $\mathrm{N_2O_5}$. For $\mathrm{Cl^-}$, the point-charge is reduced to $-0.85e$ for effective consideration of the polarization effect of water\cite{leontyev2011accounting}. The reduced charge on the ion reproduces the reversible work profile of $\mathrm{Cl^-}$ computed from the polarizable water environment (SWM4-NDP)\cite{ou2013spherical}. Each diabatic potential energy surface is shifted to match the potential energy difference between states.

\begin{figure}[t]
\centering
\includegraphics[width=0.5\textwidth]{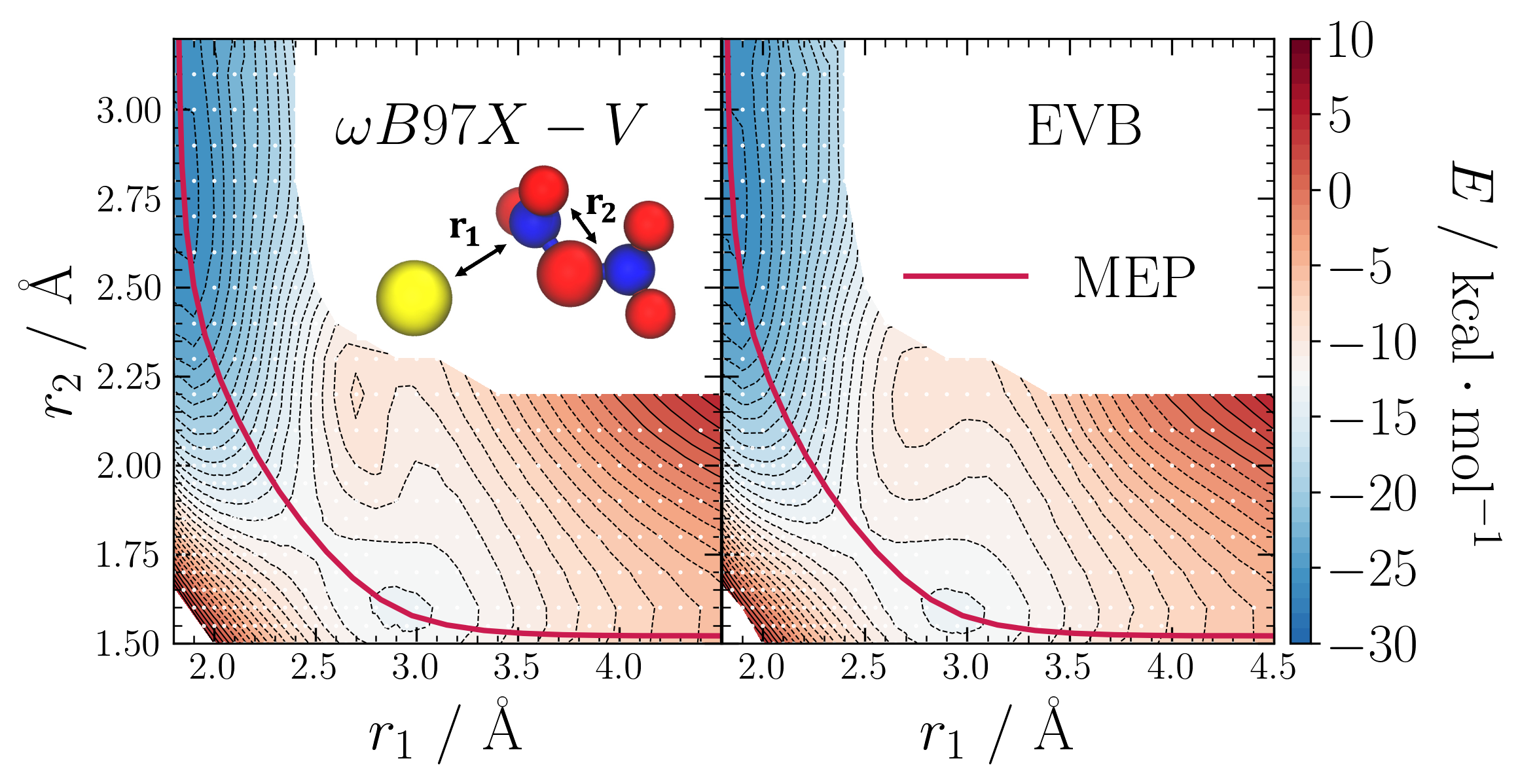}
\caption{The \textit{ab-initio} PES on $(r_1, \ r_2)$ space from geometry optimization with $\omega$B97X--V/DEF2-TZVPD (left) and EVB (right). Red solid line indicates the minimum energy path of the reaction. }
\label{fig:S1}
\end{figure}

The off-diagonal element is modeled as a function of $r_1$ and $r_2$. Accordingly, the adiabatic potential energy surface is obtained by gas-phase constrained optimizations in $(r_1, r_2)$ space. This simplification let us to devise an iterative algorithm to determine the diabatic coupling. The value of the diabatic coupling only affects the mixing coefficients of two diabatic states if $r_1$ and $r_2$ are fixed, as observed from, 
\begin{align}
  F_{ad} = \frac{\partial H_{ad}}{\partial H_{11}} F_{11} + \frac{\partial H_{ad}}{\partial H_{22}} F_{22} + \frac{\partial H_{ad}}{\partial H_{12}} F_{12} \, ,
\end{align}
where $F_{12}=0$ if $r_1$ and $r_2$ are fixed. Since a test value of $H_{12}$ determines coefficients for $F_{11}$ and $F_{22}$, geometry optimization can be performed with the empirical valence bond model and we obtained diabatic energies that provide better estimates of $H_{12}$ through $H_{12}^2 = (H_{ad} - H_{11})(H_{ad}-H_{22})$. The diabatic coupling function on $(r_1, r_2)$  is then fitted by a linear combination of 10 Gaussian functions\cite{schlegel2006empirical} centered around the transition state. The width, centers and coefficients of the Gaussian are optimized by Monte Carlo sampling using the mean squared error as a loss function. The \textit{ab-initio} and parameterized surface are plotted in Fig.~\ref{fig:S1} and show good agreement.

\begin{figure}[b]
\centering
\includegraphics[width=8.5cm]{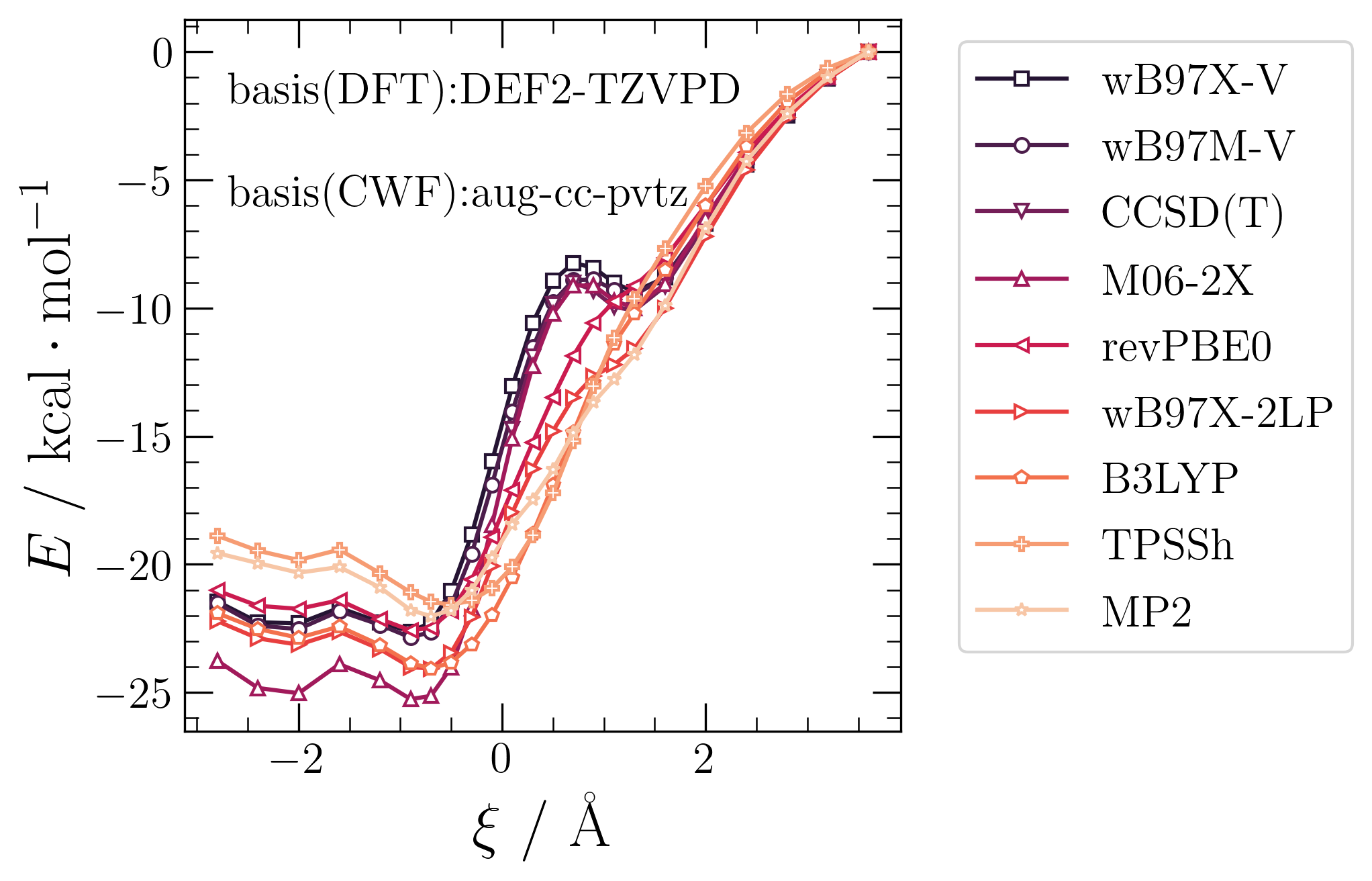}
\caption{Single-point potential energies on the structure of the minimum energy path of the reaction, evaluated with various DFT functionals and correlated wavefunction methods. Basis set for DFT and correlated wavefunction (CWF) methods are provided. }
\label{fig:S2}
\end{figure}

The transition state of the reaction is well captured on the $(r_1, \ r_2)$ space. The transition state was located by applying the freezing string method\cite{behn2011efficient}, which connects the geometry optimized ion-dipole complex of the reactant and product states. The minimum energy path was found by intrinsic reaction coordinate method\cite{fukui1970formulation} and we confirmed that all points in minimum energy path correspond unique locations on $(r_1, \  r_2)$. The performance of several density functionals and correlated wave function methods are tested by performing single-point calculations on minimum energy path geometries, which is shown in Fig.~\ref{fig:S2}. Our functional represents similar trend as CCSD(T)/aug-cc-pvtz and hybrid meta-GGA functionals.

\section{Details on molecular simulations and enhanced sampling}
We performed molecular dynamics simulations using an extended air-water interface system. A rectangular cuboid box of the size $24.74 \mathrm{\AA} \times 24.74 \mathrm{\AA} \times 100\mathrm{\AA}$, was filled with 512 water molecules and a single $\mathrm{N_2O_5}$ and $\mathrm{Cl^-}$, creating an air-water interface normal to the $z$ direction. Periodic boundary conditions are applied in all directions. The Lennard-Jones interaction are tuncated and shifted at $12\mathrm{\AA}$. Ewald summation method is applied to compute long-ranged Coulomb interaction. The equation of motion is integrated by velocity-Verlet algorithm with $0.5\mathrm{fs}$ time step. Langevin thermostat is applied to fix the temperature at $298\mathrm{K}$ with damping constant of $2\mathrm{ps}$. All molecular dynamics simulations were performed in LAMMPS\cite{thompson2022lammps} with the in-house code.

The 1-dimensional free energy profiles in Fig.~\ref{fig:1}b are computed from umbrella sampling along $z$. The density profile of water is decayed to a half of its bulk value at $z=0\mathrm{\AA}$, positive and negative $z$ correspond to the gas and the solution side, respectively. We prepared 58 independent simulations for each species with harmonic biasing potentials, 
\begin{align}
    U(z) =\frac{1}{2} k (z-z_c^i)^2 \, ,
\end{align}
where $k=8\mathrm{kcal\cdot mol^{-1}} \mathrm{\AA}^{-2}$ and $z_c^i$, the center--of--mass location of species $i$, are ranging from $-12.1\mathrm{\AA}$ to $10.7\mathrm{\AA}$ equally spaced by $0.4\mathrm{\AA}$. For each simulation, we ran $2.7\mathrm{ns}$ of production simulation after $0.3\mathrm{ns}$ of equilibration simulation. Sampled histograms of $z$ are merged together with the weighted histogram analysis method (WHAM)\cite{kumar1995multidimensional} and errors of the free energy are estimated by running 100 bootstrapping trials\cite{grossfield2011wham}.

The 2-dimensional umbrella sampling used for the free energy surface in Fig.~\ref{fig:2}a were performed by applying two harmonic potentials,
\begin{align}
    U(\xi, z) = \frac{1}{2} k_{\xi} (\xi-\xi_0)^2 + \frac{1}{2} k_z (z-z_0)^2 \, ,
\end{align}
where $k_{\xi}=200\mathrm{kcal\cdot mol^{-1}} \mathrm{\AA}^{-2}$, $k_z=20\mathrm{kcal\cdot mol^{-1}} \mathrm{\AA}^{-2}$, $\xi_0$ values are ranging from $-0.08\mathrm{\AA}$ to $5.62\mathrm{\AA}$ equally spaced by $0.06\mathrm{\AA}$, and $z_0$ values are ranging from $-7\mathrm{\AA}$ to $5\mathrm{\AA}$ equally spaced by $0.5\mathrm{\AA}$. In the vicinity of the transition state, we performed additional simulations with $k_{\xi}=400\mathrm{kcal\cdot mol^{-1}} \mathrm{\AA}^{-2}$ and $\xi_0$ values ranging from $0.31\mathrm{\AA}$ to $0.58\mathrm{\AA}$ equally spaced by $0.03\mathrm{\AA}$. A total of 2650 independent simulations are performed for $1.8\mathrm{ns}$ for the production simulation after $0.2\mathrm{ns}$ of equilibration.

The transmission coefficients in Fig.~\ref{fig:2}c are computed from an ensemble of trajectories started from the transition state at three fixed $z$ locations; $z=-12.3\mathrm{\AA}$ for the bulk, $z=0\mathrm{\AA}$ for the interface, and $z=5\mathrm{\AA}$ for the gas case. We sampled 300 initial configurations for each case by running constrained simulation at the transition state. For each configuration, we saved $\xi$ and $\Delta s$ values for analysis. Then, we generated 100 different Maxwell-Boltzmann velocities for each configuration and propagated the system for $0.5\mathrm{ps}$ to let them relax to either reactant or product basin. Moreover, we repeated the propagation with inverted velocities to improve the convergence of the transmission coefficient using the symmetry,
\begin{align}
    \kappa(z,t) = \frac{\left< v(0) h_P(\xi(t)) \right>^{*}_{z} }{\left<|v|\right>/2} = -\frac{\left< v(0) h_R(\xi(t)) \right>^{*}_{z} }{\left<|v|\right>/2} \, ,
\end{align}
where the indicator function for the reactant ($h_R$) and product ($h_P$) states are step-functions which have the kink at initial $\xi$ values, i.e.,
\begin{align}
h_P(\xi) =
\begin{cases}
    1 & \text{if } \xi \le \xi^*, \\
    0 & \text{otherwise} \, ,
\end{cases}
\end{align}
where $\xi^*$ is the transition state value corresponding to the $z$ constraint and $h_R = 1- h_P$. The committor function in Fig.~\ref{fig:2}d, $P_B (\xi, \Delta s)$, is defined as the conditional probability of reaching to the product state at $t=0.5\mathrm{ps}$ given the initial configuration's $\xi$ and $\Delta s$ values.

\section{Heterogeneous Bimolecular Rate Expression}
The rate is defined as the time derivative of site-site correlation function,
\begin{align}
    k(t) =\frac{\partial_t \left<h_R(0) h_P(t)\right> }{\left<h_R\right>} \label{eqn:flux} \, ,
\end{align}
where $h_R$ and $h_P$ are indicator functions of the reactant and the product state. The expression provides the phenomenological rate constant at time $t$ much longer than the molecular relaxation time scale but still shorter than the typical barrier-crossing time scale. The transition state theory (TST) rate is the asymptotic limit of $t\rightarrow 0^+$ of the expression. By introducing step-function-like indicator functions with $\xi>\xi^*$ as the reactant domain and $\xi < \xi^*$ as the product domain, the TST rate becomes,
\begin{align}
     k^\mathrm{TST} =  \frac{\left<\dot{\xi} \theta(\dot{\xi}) \delta(\xi-\xi^*) \right> }{\left< \theta(\xi-\xi^*) \right> } \, ,  
\end{align}
where $\theta$ is Heaviside step function and $\delta$ is Dirac delta function. The velocity of $\xi$ can be integrated out by,
\begin{align}
    \left<\dot{\xi} \theta(\dot{\xi}) \delta(\xi-\xi^*) \right> =  \left< \sqrt{ \frac{\kB T|\mathbf{G}|}{2\pi}} \delta(\xi-\xi^*)\right> \, ,
\end{align}
where the inverse effective mass $|\mathbf{G}|$ is defined as,
\begin{align}
    |\mathbf{G}| = \sum_i \frac{1}{m_i}\left|\frac{\partial \xi }{\partial R_i} \right|^2 = \frac{1}{m_{\mathrm{Cl}}} + \frac{1}{m_{\mathrm{NO_3}}} + \frac{2(1-\cos\theta)}{m_{\mathrm{NO_2}}} \, ,
\end{align}
where $R_i$ is the Cartesian coordinate vector and $m_i$ is the mass of particle $i$, and  $\theta$ is the angle between $r_1$ and $r_2$ vectors. Around the transition state, $\theta$ is almost a right angle, so we considered the inverse mass as a constant. By introducing the free energy profile along $\xi$, $F(\xi)$, the TST rate becomes,
\begin{align}
     k^\mathrm{TST} = \sqrt{ \frac{\kB T|\mathbf{G}|}{2\pi} } \frac{\exp(-F(\xi^*)/\kB T)}{\int_{\xi^*}^{\infty} d\xi \exp(-F(\xi)/\kB T)} .\label{eqn:tstrate}
\end{align}
By introducing the asymptotic scaling of the free energy, $F(\xi) = F(\xi^a) - 2 \kB T \ln \left(r_1/r_1^{a} \right)$, which is valid for large separation of species with no correlation, the denominator can be integrated by,
\begin{align}
    \int_{\xi^*}^{\infty} d\xi \exp(- F(\xi)&/\kB T)  = \int_{\xi^*}^{\xi^a} d\xi \exp(- F(\xi)/\kB T) \notag \\
    & +\exp(-F(\xi^a)/\kB T) \frac{V}{4\pi (r_1^a)^2} \, ,
\end{align}
where $V$ is the macroscopic volume of the system. The second term dominates the integral unless the reactant well is deep enough so that the reaction becomes almost unimolecular. The bimolecular reactive flux in Eq.~\ref{eqn:flux} satisfies,
\begin{align}
    \partial_t N_{\mathrm{Cl}} = -\kappa k^\mathrm{TST} N_{\mathrm{N}} N_{\mathrm{Cl}} \, ,
\end{align}
where $N_\mathrm{Cl}$ and $N_\mathrm{N}$ is the number of $\mathrm{Cl^-}$ and $\mathrm{N_2O_5}$, respectively, and $\kappa$ is the transmission coefficient. Dividing both sides with the system's volume, we get,
\begin{align}
    \partial_t \rho_\mathrm{Cl} = -(\kappa k^\mathrm{TST} V) \rho_\mathrm{Cl} \rho_\mathrm{N} = -\kappa k_\mathrm{C}^\mathrm{TST} \rho_\mathrm{Cl} \rho_\mathrm{N} \, ,
\end{align}
where $\rho_\mathrm{Cl}=N_\mathrm{Cl}/V$, $\rho_\mathrm{N}=N_\mathrm{N}/V$, and $k^\mathrm{TST}_\mathrm{C}$ is the bimolecular rate for the chlorination reaction, which is,
\begin{align}
    k^\mathrm{TST}_\mathrm{C} = \left[ \sqrt{ \frac{\kB T|\mathbf{G}|}{2\pi} }  \frac{3}{r_1^a} e^{-\Delta F/\kB T} \right] \frac{4\pi (r_1^a)^3}{3} \, , \label{eqn:bmolrate}
\end{align}
where $\Delta F = F(\xi^*) - F(\xi^a)$ is the free energy barrier.

The heterogeneous bimolecular rate is derived by applying constraints. For the system with azimuthal symmetry in $z$ direction, the free energy barrier is conditioned on the $\mathrm{Cl^-}$ $z$ location. Accordingly, 2-dimensional free energy in Fig.~\ref{fig:2}a replaces $F(\xi)$. Moreover, we incorporate the local density enhancement of $\mathrm{N_2O_5}$ by introducing $\nu(z)$ in Eq.~\ref{eqn:Lz}. This is necessary since $\mathrm{N_2O_5}$ position is not constrained; the phenomenological reactive flux at $z$ depends on density profiles of both species at $z$, while the barrier-crossing frequency is proportional to the amount of $\mathrm{N_2O_5}$ molecules inside the sphere of radius $r_1^a$. The final expression provided in Eq.~\ref{eqn:heterorate} satisfies,
\begin{align}
    \partial_t \rho_{\mathrm{N}}(z, t)  = - \kappa (z)  k_\mathrm{C}^\mathrm{TST}(z) \rho_{\mathrm{N}}(z, t) \rho_{\mathrm{Cl}}(z, t) \, ,
\end{align}
where density profiles depend both on position and time.

\section{Reaction-Diffusion Equation}
The heterogeneous reactive uptake process is modeled with Eq.~\ref{eqn:RDE}. Given the sticking probability of $1$, the fate of initially adsorbed $\mathrm{N_2O_5}$ at the air-water interface determines the reactive uptake coefficient as in Eq.~\ref{eqn:uptake}. The density profile of $\mathrm{N_2O_5}$ is evolved through the overdamped Fokker-Planck equation with additional reactive loss term, which is valid for condensed phase diffusion. Instead of using underdamped Fokker-Planck dynamics for the evaporation process, we placed the absorbing boundary condition at the gas side of the interface. Reflecting boundary condition is applied far away from the interface relative to the reaction-diffusion length of the hydrolysis reaction, $l_{RD} = \sqrt{D/k_H} = 32.4\mathrm{nm}$. 

The reaction-diffusion equation is propagated by the finite difference method. The density profile $\rho_\mathrm{N}(z, t)$ follows,
\begin{align}
    \partial_t \rho_\mathrm{N} = &-\partial_z (F_\mathrm{N}(z)\gamma^{-1}(z) \rho_\mathrm{N} )  + \partial_z ( D(z) \partial_z \rho_\mathrm{N}) \notag \\
    &- k_\mathrm{T}(z)\rho_\mathrm{N} \, ,
\end{align}
where $\gamma^{-1}(z) = \beta D(z)$. By using $i$ and $n$ for the discrete index for $z$ and $t$, respectively, we get
\begin{align}
    &\frac{\rho_i^{n+1} - \rho_i^{n}}{\Delta t}  = -\frac{ F_{i+1} \gamma_{i+1}^{-1} \rho_{i+1}^{n} - F_{i-1} \gamma_{i-1}^{-1} \rho_{i-1}^{n}}{2\Delta z} \\ \notag 
    &+ \frac{D_{i+1/2} (\rho_{i+1}^{n} - \rho_{i}^{n}) - D_{i-1/2} (\rho_{i}^{n} - \rho_{i-1}^{n}) }{(\Delta z)^2} -k_\mathrm{T}(z_i)\rho^n_i \, ,
\end{align}
where $\rho_i^n = \rho_\mathrm{N}(z_i, t_n)$, $F_i = F_N(z_i)$, $\gamma_i^{-1} = \gamma^{-1}(z_i) $, $D_i = D(z_i)$ and $D_{i\pm 1/2} = (D_i + D_{i\pm 1})/2$. We set $\Delta t= 18\mathrm{fs}$ and $\Delta z = 0.2\mathrm{\AA}$. We descritized the marginal free energy profile of $\mathrm{N_2O_5}$ in Fig.~\ref{fig:1}b by fitting it with a smooth analytic function. We assigned two domains by introducing $z_B=-5\mathrm{\AA}$ as a boundary; the interface domain as $z_{ABC} > z > z_B$ and the bulk domain as $z_{RBC}<z< z_B$. We modeled the position-dependent diffusion coefficient and hydrolysis rate constant as step functions with kinks at $z=z_B$. Diffusion coefficients are set to $D_B = 1.89\mathrm{nm^2/ns}$ for the bulk domain, and $D_I = 5.30\mathrm{nm^2/ns}$ for the interface domain. 

The location of the absorbing boundary is determined from the detailed balance relation between evaporation flux and adsorption flux,
\begin{align}
    k_e \int_{z_B}^{z_{ABC}} dz \rho^{eq}_{\mathrm{N}}(z) = S \frac{\left<|\bold{v}|\right>}{4} \rho_g = S\sqrt{\frac{\kB T}{2\pi m}} \rho_g \, , \label{eqn:DB}
\end{align}
where $m$, $\bold{v}$ and $\rho_g$ are the mass, the velocity vector and the gas-phase density of $\mathrm{N_2O_5}$, respectively. Since the equilibrium density profile is related to the free energy profile as, $\rho^{eq}_\mathrm{N} (z)/\rho_g = \exp(-(F_\mathrm{N}(z)-F_\mathrm{N} (\infty)/\kB T)$, the evaporation rate can be evaluated if $z_{ABC}$ is determined from the decay of the density profile. 
When the absorbing boundary is located at $z_{ABC}=6.58\mathrm{\AA}$, both expressions result in the same evaporation rate of $k_e=0.298\mathrm{/ns}$.

For the system with pure hydrolysis reaction, the 
reported value of the uptake coefficient from pure water droplet is reproduced when the bulk hydrolysis rate is $k_{B,H}= 0.4/\mathrm{\mu s}$, regardless of the interfacial hydrolysis rate, $k_{I, H}$. Since the uptake process is dominated by the bulk reaction, we tested the performance of the resistor model. The reactive uptake coefficient $\gamma$ of the bulk-dominant reaction at the micro-sized aerosol is given in terms of the mass accommodation coefficient ($\alpha$) and the resistance of the bulk reactivity ($\Gamma_{rxn}$), 
\begin{align}
    &\frac{1}{\gamma} = \frac{1}{\alpha} + \frac{1}{\Gamma_{rxn}}, \notag \\ 
    \frac{1}{\alpha} = \frac{k_{s} + k_{e}}{S k_{s}}, &\indent \frac{1}{\Gamma_{rxn}} = \frac{\overline{v}}{4H\sqrt{Dk_{B, H}}} \, , \label{eqn:resistor}
\end{align}
where $S$ is the sticking probability, $D$ is the bulk diffusion coefficient, $H$ is Henry's law constant, $\overline{v}$ is the mean molecular velocity, and $k_{s}$, $k_{e}$, $k_{B, H}$ are first-order rates for solvation, evaporation and bulk hydrolysis reaction, respectively. Although the solvation process does not follow the first-order kinetic law due to the coupled diffusive motion, we defined it as an inverse of the first-passage time to the boundary of the interface domain $z_B$, which is $k_{s}=2.68 \ /\mathrm{ns}$. This choice would not affect to the uptake in pure hydrolysis case, as the bulk resistance dominates the uptake process and $\gamma$ is insensitive to its choice. However, it turns out as an optimal choice if we matched the solution of Eq.~\ref{eqn:RDE} with the resistor model in bulk dominant case.

\begin{figure}[t]
\centering
\includegraphics[width=8.5cm]{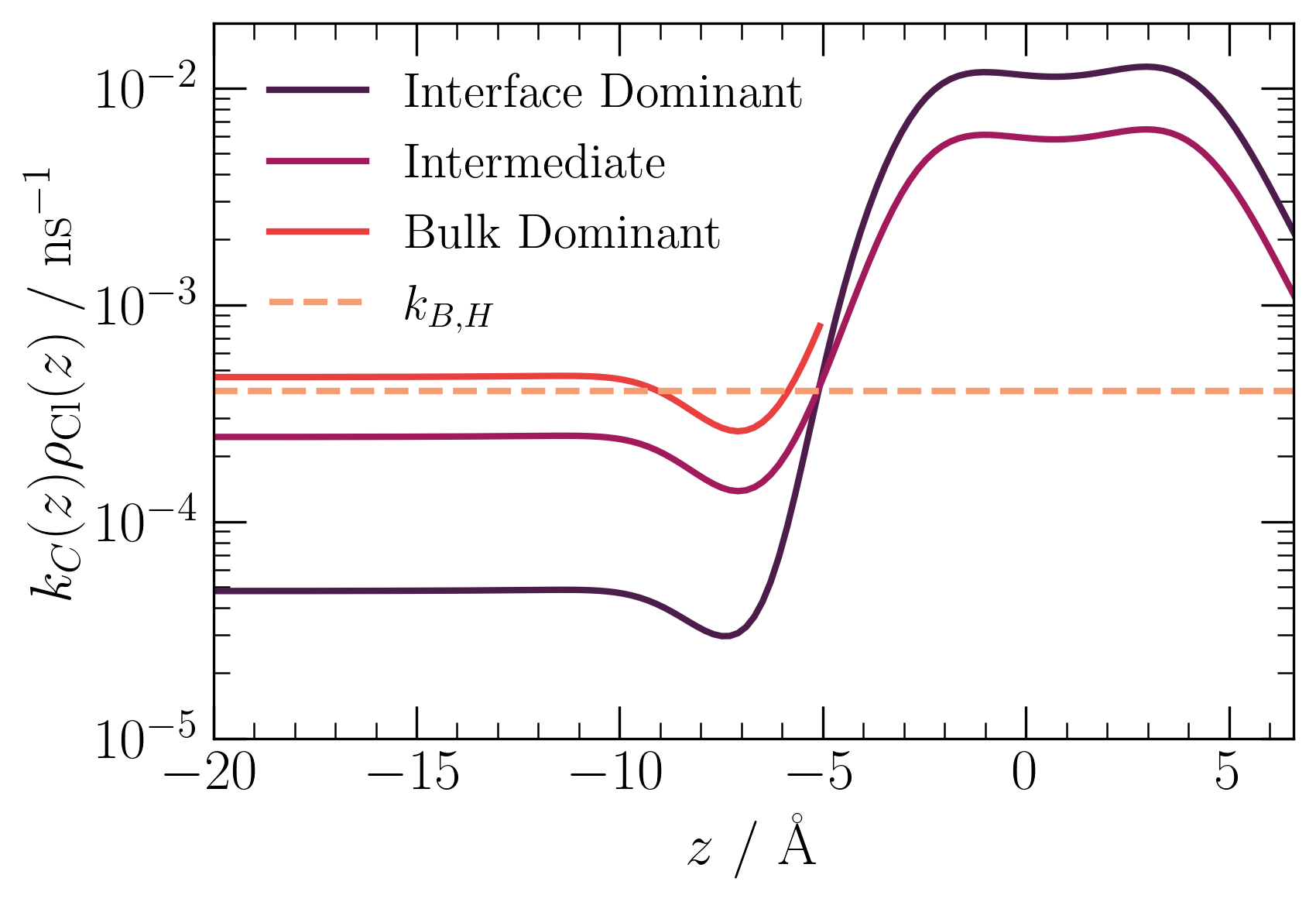}
\caption{ Shifted pseudo chlorination rate profiles of three scenarios. For facile comparison, $\bar{\rho}_{\mathrm{Cl}}$ is set to $0.05$ M, where the branching ratio becomes near half. Each profile is shifted to reproduce measured branching ratio at $\bar{\rho}_{\mathrm{Cl}} =0.0054\mathrm{M}$. The interface dominant hypothetical profile is obtained by multiplying $9\tanh(z+5)/2 + 11/2$ to the intermediate profile. For the bulk dominant case, interfacial rate is set to 0. }
\label{fig:S3}
\end{figure}
\begin{figure}[b]
\centering
\includegraphics[width=0.45\textwidth]{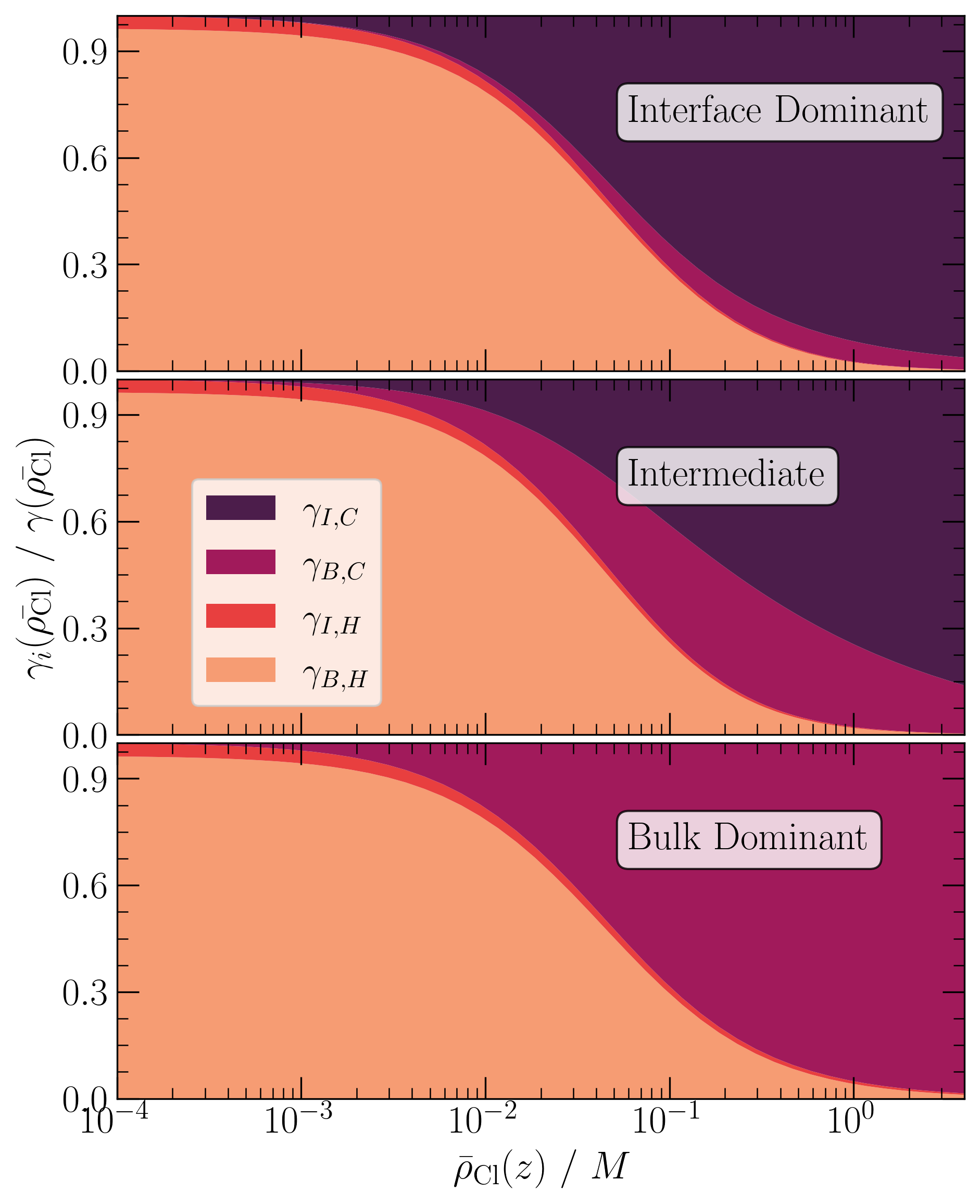}
\caption{ Relative reactive uptake coefficient for each reactive channel, $\gamma_i$, where $i=$I,C (interfacial chlorination), B,C (bulk chlorination), I,H (interfacial hydrolysis), B,H (bulk hydrolysis)}
\label{fig:S5}
\end{figure}

The second reaction channel due to the chlorination provides additional sink to the $\mathrm{N_2O_5}$ density loss, which in general depends on the time-dependent density profile of $\mathrm{Cl^-}$. However, when it comes to a trace amount of gas phase $\mathrm{N_2O_5}$ compared to the amount of $\mathrm{Cl^-}$ at sea-spray aerosol, the depletion of $\mathrm{Cl^-}$ density is negligible. Therefore, we solved Eq.~\ref{eqn:RDE} with additional sink term proportional to the pseudo-first-order rate, $k_{\mathrm{C}}(z) \rho_{\mathrm{Cl}}(z)$. The pseudo rate are shifted to reproduce experimentally-measured branching ratio of $\Phi = 0.108$ at $\bar{\rho_{\mathrm{Cl}}}=0.0054\mathrm{M}$. Two additional hypothetical cases are considered and the rate profiles used for three cases are summarized in Fig.~\ref{fig:S3}. Moreover, relative losses through either hydrolysis or chlorination, at either interface or bulk are summarized in Fig.~\ref{fig:S5}. The enhanced interfacial chlorination rate takes up large portions of total loss even at moderate bulk $\mathrm{Cl^-}$ concentration, which emphasizes the importance of considering interfacial reactive flux to the analysis of uptake process. 

\section{New Analytic Expression for Reactive Uptake}
An intuition for the generalization of the resistor model is achieved from the observation that both the evaporation process and interfacial reactions follow the first-order kinetics. In other words, interfacial reactivity can be absorbed into the enhanced evaporation rate at the interface. Additionally, enhanced evaporation rate lowers the solubility, which reduces Henry's law constant. Therefore, we introduce modifications to the mass accommodation coefficient $\alpha'$ and Henry's law constant $H'$,
\begin{align}
    \alpha' = S \frac{k_{s}}{k_{s} + k_{e} +k_\mathrm{I, T}}, \indent H' = H \frac{k_{e}}{k_{e} + k_\mathrm{I,T}} \, ,
\end{align}
where $k_\mathrm{I, T}$ is the total interfacial reaction rate. With this modification, we applied the traditional bulk resistor model to evaluate the bulk contribution to the uptake coefficient, $\gamma_\mathrm{B}$,
\begin{align}
    \gamma_\mathrm{B} = \left(\frac{1}{\alpha^{'}} + \frac{\overline{v}}{4H'\sqrt{Dk_\mathrm{B, T}}} \right)^{-1} \, , \label{eqn:gammabulk}
\end{align}
where $k_\mathrm{B,T}$ is the total bulk reaction rate. The $\gamma_\mathrm{B}$ provides the reactive loss due to the bulk reactivity, while we need to account the interfacial loss to evaluate the total uptake coefficient. Note that $(1-\gamma_\mathrm{B})$ is the loss through either evaporation or interfacial reactions in our reaction-diffusion model. They are competing each other at the interface, but we can weight them through their rate constant. Therefore, the total reactive uptake coefficient becomes,
\begin{align}
    \gamma_\mathrm{T} = \gamma_\mathrm{B} + (1-\gamma_\mathrm{B}) \frac{k_\mathrm{I,T}}{k_e + k_\mathrm{I,T}} \equiv \gamma_\mathrm{B} + (1-\gamma_\mathrm{B})\gamma_\mathrm{I} \, , \label{eqn:gammainterface}
\end{align}
where we introduced the ideal interfacial uptake coefficient, $\gamma_\mathrm{I}$, which is the total uptake coefficient if the evaporation is so fast that diffusion into bulk becomes irrelevant. 

\begin{table}
\caption{\label{tab:table1}%
The chlorination rates for three scenarios used as parameters for the new analytic expression. Units for the bimolecular rate are $\mathrm{\mu s ^{-1} \cdot M^{-1}}$.
}
\begin{ruledtabular}
\begin{tabular}{lrr} \label{table:1}
\textrm{}&
$k_\mathrm{B, C} $&
\textrm{$k_\mathrm{I,C}$}\\
\colrule
Interface Dominant & 0.9844 & 231.0 \\
Intermediate & 5.058 & 118.7 \\
Bulk Dominant & 9.557 & 0.0 \\
\end{tabular}
\end{ruledtabular}
\end{table}

The analytic expressions we just introduced can be straightforwardly generalized to multiple reaction channel problems. For example, if we consider the hydrolysis ($k_\mathrm{H}$) and the chlorination ($k_\mathrm{C}$), the reaction rate at the interface $k_\mathrm{I,T} = k_\mathrm{I,H} + k_\mathrm{I,C}\bar{\rho}_\mathrm{Cl}$, and at the bulk $k_\mathrm{B,T} = k_\mathrm{B,H} + k_\mathrm{B,C}\bar{\rho}_\mathrm{Cl}$ provide expressions for the total uptake coefficient from Eq.~\ref{eqn:gammabulk} and \ref{eqn:gammainterface}. Moreover, the uptake coefficient due to each reaction channel, $\gamma_\mathrm{H}$ and $\gamma_\mathrm{C}$, can be obtained by simple reweighting, 
\begin{align}
    \gamma_\mathrm{H} &= \gamma_\mathrm{B} \frac{k_\mathrm{B,H}}{k_\mathrm{B,H}+k_\mathrm{B,C}\bar{\rho}_\mathrm{Cl} }  + (1-\gamma_\mathrm{B}) \gamma_\mathrm{I} \frac{k_\mathrm{I,H}}{k_\mathrm{I,H}+k_\mathrm{I,C}\bar{\rho}_\mathrm{Cl}} \, , \notag \\
    \gamma_\mathrm{C} &= \gamma_\mathrm{B} \frac{k_\mathrm{B,C}\bar{\rho}_\mathrm{Cl}}{k_\mathrm{B,H}+k_\mathrm{B,C}\bar{\rho}_\mathrm{Cl}}  + (1-\gamma_\mathrm{B}) \gamma_\mathrm{I} \frac{k_\mathrm{I,C}\bar{\rho}_\mathrm{Cl}}{k_\mathrm{I,H}+k_\mathrm{I,C}\bar{\rho}_\mathrm{Cl}}.
\end{align}
The branching ratio $\Phi = \gamma_C / (\gamma_\mathrm{H} + \gamma_\mathrm{C})$ and the total uptake coefficient $\gamma_\mathrm{T} = \gamma_\mathrm{C} + \gamma_\mathrm{H}$ are plotted in Fig.~\ref{fig:3}b and c, respectively. Parameters we used for the analytic model are summarized in Table~\ref{table:1} and here. $D = 1.89\mathrm{nm^2/ns}$, $H=76.93$, $\bar{v}=241.66\mathrm{nm/ns}$, $k_e=0.298 / \mathrm{ns}$, $k_s=2.680/\mathrm{ns}$, $S=1$, $k_\mathrm{B,H} = k_\mathrm{I,H} = 0.4 \mathrm{\mu s^{-1}}$. 
\end{document}